\documentclass[12pt]{article}
\usepackage[dvips]{graphics,color}
\usepackage{psfrag}
\usepackage{psfig,epsfig}
\usepackage{graphicx}
\usepackage{amssymb}
\usepackage{amsmath}
\usepackage{epsf}
\usepackage[parfill]{parskip}
\input{epsf}
\newlength{\extraspace}
\newlength{\extraspaces}
\makeatletter\@addtoreset{equation}{section}\makeatother

\newcommand{\ba}{\begin{eqnarray}
\addtolength{\abovedisplayskip}{\extraspaces}
\addtolength{\belowdisplayskip}{\extraspaces}
\addtolength{\abovedisplayshortskip}{\extraspace}
\addtolength{\belowdisplayshortskip}{\extraspace}}

\newcommand{\Tr}{{\rm Tr\,}}
\newcommand{\tr}{{\rm tr\,}}
\newcommand{\ea}{\end{eqnarray}}

\newcommand{\IZ}{\mathbb{Z}}
\newcommand{\IC}{\mathbb{C}}
\newcommand{\IP}{\mathbb{P}}

\newcommand{\IR}{\mathbb{R}}

\newcommand{\re}{{\rm Re \,}}

\def \nn{\nonumber}

\newcommand{\cO}{{\cal O}}

\def \min{{\rm min}}
\def \hkappa{\kappa \!\!\!/\,}
\def \bx{\bar{x}}

\def \bk{{\bf k}}
\def \bx{{\bf x}}
\def \la{\langle}
\def \ra{\rangle}
\def \blb{\{ \!\!\{ \!\!\!\{}
\def \brb{\} \!\!\} \!\!\!\}}

\newdimen\tableauside\tableauside=1.0ex
\newdimen\tableaurule\tableaurule=0.4pt
\newdimen\tableaustep
\def\phantomhrule#1{\hbox{\vbox to0pt{\hrule height\tableaurule width#1\vss}}}
\def\phantomvrule#1{\vbox{\hbox to0pt{\vrule width\tableaurule height#1\hss}}}
\def\sqr{\vbox{%
  \phantomhrule\tableaustep
  \hbox{\phantomvrule\tableaustep\kern\tableaustep\phantomvrule\tableaustep}%
  \hbox{\vbox{\phantomhrule\tableauside}\kern-\tableaurule}}}
\def\squares#1{\hbox{\count0=#1\noindent\loop\sqr
  \advance\count0 by-1 \ifnum\count0>0\repeat}}
\def\tableau#1{\vcenter{\offinterlineskip
  \tableaustep=\tableauside\advance\tableaustep by-\tableaurule
  \kern\normallineskip\hbox
    {\kern\normallineskip\vbox
      {\gettableau#1 0 }%
     \kern\normallineskip\kern\tableaurule}%
  \kern\normallineskip\kern\tableaurule}}
\def\gettableau#1 {\ifnum#1=0\let\next=\null\else
  \squares{#1}\let\next=\gettableau\fi\next}

\begin{document}
\noindent
\begin{titlepage}

\begin{center}
\hfill ITFA-2006-23\\
\hfill hep-th/0606112\\ 
\vskip 1cm {\LARGE 
The Wave Function Behavior \\ \vspace{0.03cm}of the \\ \vspace{0.03cm}Open Topological String Partition Function \\  \vspace{0.2cm}on the Conifold}
\vskip 2cm {Amir-Kian Kashani-Poor}\\ \vskip 0.5cm

{\it Institute for Theoretical Physics, University of Amsterdam\\
1018 XE Amsterdam, The Netherlands\\}

\end{center} 
\vskip 1cm
\begin{abstract}
We calculate the topological string partition function to all genus on the conifold, in the presence of branes. We demonstrate that 
the partition functions for different brane backgrounds (smoothly connected along a quantum corrected moduli space) can be interpreted as the same wave function in different polarizations. This behavior has a natural interpretation in the Chern-Simons target space description of the topological theory. Our detailed analysis however indicates that non-perturbatively, a modification of real Chern-Simons theory is required to capture the correct target space theory of the topological string. 

We perform our calculations in the framework of a free fermion representation of the open topological string, demonstrating that this framework extends beyond the simple $\IC^3$ geometry. The notion of a fermionic brane creation operator arises in this setting, and we study to what extent the wave function properties of the partition function can be extended to this operator. 
\end{abstract}
\end{titlepage}
\newpage

\section{Introduction}
Just as in physical string theory, topological string amplitudes were originally  defined genus by genus from a worldsheet formulation. Much progress has been made in the meantime in understanding these theories from a target space point of view, which often provides better computability than the worldsheet approach \cite{BCOV2, allgenus, DV}. Unlike the worldsheet description, the target space approach should in principle provide a non-perturbative definition of the theory. In this paper, we will study a manifestation of the underlying target space description in the purely perturbative (albeit all genus) partition function of the open topological string: given a path integral on a space with boundary, the partition function defines a wave function with phase space the restriction of field space to the boundary. For a path integral on a non-compact space, we can consider the boundary at infinity. In the Hamiltonian formalism, we must choose a polarization on field space, and only fix half of the fields on the boundary. The claim that the partition function defines a wave function is really the statement that it transforms as a wave function under a change of this polarization. This transformation property of the partition function is a central focus of this paper.
The question we address, in the example of the resolved conifold, is how the choice of polarization manifests itself in the open topological string partition function. The intriguing observation of \cite{ADKMV} for the case of the vertex, i.e. the geometry $\IC^3$, is that different brane placements in the geometry appear to correspond to different polarizations. Consequently, partition functions associated to these different brane placements should be related by canonical transformations. In this paper, we study this claim for the case of the conifold. We find that the proper choice of canonical coordinates depends on such data as the distinction between branes and antibranes, and the choice of K\"ahler cone. We emphasize the unusual form of phase space, which with a symplectic form of type (2,0) presents a holomorphic generalization of the conventional setup.
We verify that partition functions for different brane placements are indeed related by canonical transformations, upon choosing appropriate complex integration contours, and neglecting terms of order ${\cO(e^{-1/g_s})}$. Our findings appear to point towards the need for a holomorphic version of Chern-Simons theory (not to be confused with holomorphic Chern-Simons theory, which is 6 dimensional, or Chern-Simons theory for a complex gauge group), which coincides with conventional Chern-Simons theory perturbatively. 

The authors of \cite{ADKMV} propose going beyond the picture of partition functions transforming as wave functions: they introduce brane creation operators using a free fermion formulation of the topological vertex, and assign the transformation properties to these operators. Studying the free fermion formulation presents the second thematic thrust of this paper. We verify a conjecture of \cite{ADKMV} about the form of the closed vacuum state $| Z \rangle$  (that it be presentable as the exponential of a sum of fermionic bilinears acting on the vacuum) for the conifold. While we have difficulty with the physical interpretation of multiple fermionic operators as creating multiple branes, we press ahead to study their $n$-point functions in their own right. The conjecture of \cite{ADKMV} that these operators can be transformed into each other via canonical transformation implies relations between various $n$-point functions. We are able to verify such a relation for a carefully selected pair of $2$-point functions on the conifold, but not in general. There is evidently much structure here that remains to be explored.

The organization of the paper is as follows. After introducing the geometric setup and the free fermion formalism in sections \ref{setup} and \ref{s:winding}, we turn to the question of encoding the open topological string data of the conifold in a state $|Z\rangle$ given by the exponential of a sum of fermion bilinears in section \ref{s:bilinear}. In section \ref{s:fermionops}, we introduce the fermion operator of \cite{ADKMV}, study its interpretation in terms of the A-model (\ref{s:oneandtwo}), and calculate the one and two point function of the operator for the conifold (\ref{s:eval12}). The one point function is the partition function of the open topological string in the presence of a single brane, and the main object of study of section \ref{s:wave}.\footnote{The route via the one point function is not the fastest way to determine this partition function -- we take it merely because we are already more than halfway there after the considerations of section \ref{s:encoding}.} After motivating the connection between brane placement and polarization in section \ref{s:bpcp}, we turn to assigning canonical coordinates to different open string configuration on the conifold in section \ref{s:assignment}. The transformation properties of the partition function that follow from the assignment of canonical coordinates are studied in section \ref{s:transform}, and we find that they are satisfied only up to ${\cal O}(e^{-1/g_s})$ corrections. In section \ref{s:non-per prelim}, we present some comments regarding these corrections. Finally, in section \ref{s:transop}, we study the transformation properties of 2-point functions in one example. We end with conclusions. Calculations and some general remarks on linear canonical transformations, the B-model geometries, Frobenius' formula, basic hypergeometric series, and Schur function identities are presented in the appendices.

\section{$Z_{open}$ encoded in $|Z\rangle$} \label{s:encoding}
\subsection{The setup}  \label{setup}
We consider the A-model on local toric varieties. Our focus is the resolved conifold ${\cal O}(-1) \oplus {\cal O}(-1) \rightarrow \IP^1$, but we will always precede our analysis by performing the calculation in the case of $\IC^3$, which was considered in \cite{ADKMV}.

\begin{figure}[h]
\psfrag{x}{$|X_1|^2$}
\psfrag{y}{$|X_2|^2$}
\psfrag{z}{$|X_3|^2$}
\psfrag{w}{$|X_4|^2$}
\begin{center}
\epsfig{file=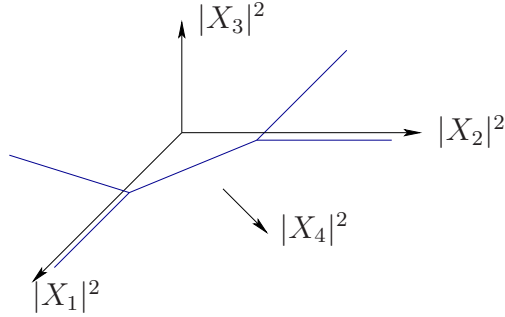, height=1.5in}
\end{center}
\caption{\small Conifold as $T^3$ fibration over manifold with corners. \label{fig:conifoldasfibr}}
\end{figure}

We will find it convenient to think of these varieties as $T^3$ fibrations over manifolds with edges and corners. This is how they arise in the gauged linear $\sigma$ model construction of Witten \cite{phases}. For $\IC^3$ in this presentation, the absolute values $|X_i|^2$ of the three complex coordinates $X_i$ on $\IC^3$ coordinatize the base of the fibration, while the phases coordinatize the $T^3$ fiber. The base hence can be thought of as an octant of the Cartesian coordinate system. On each face, edge, corner of the base, one, two, three 1-cycles respectively of the fiber degenerate. A similar presentation of the conifold is obtained via a symplectic quotient construction. We consider the zero locus
\ba \label{momentcfld}
|X_1|^2 + |X_2|^2 - |X_3|^2 - |X_4|^2 &=& t \,
\ea
in $\IC^4$, modded out by a $U(1)$ action on the phases. This equation shaves off some corners, pictorially speaking, of the octant, see figure \ref{fig:conifoldasfibr}, and again, on each face, edge, corner of the base, one, two, three 1-cycles respectively of the fiber degenerate. In the following, we will sometimes depict these geometries simply by indicating the edges and corners (or vertices) along which the fiber degenerates. We will refer to this locus as the toric skeleton.

Branes can wrap Lagrangian submanifolds in the A-model (though more general branes are possible and have recently risen to prominence). Lagrangian submanifolds in toric varieties were studied in \cite{AV}, whose analysis we briefly review here. For the treatment from a boundary $\sigma$ model point of view, see \cite{hori}. The class we will study, following \cite{AV}, is given by fibrations $F$ with generic fiber $T^2$ that project to straight lines with rational slope in the base manifold (the rationality condition is to ensure that the orbits in the $T^3$ fiber determined by the Lagrangian condition close upon themselves to yield $T^2$). As the base is 3 dimensional, such lines are given as the zero locus of two linear polynomials in the variables $|X_i|^2$, hence by two 4-tuples ${\bf q_1}, {\bf q_2}$, such that
\ba
\sum q_I^i |X_i|^2 = c_I \,. \nn
\ea
Equivalently, introduce a vector ${\bf v}$ in $\IR^4$ such that
\ba
|X_i|^2 = s v_i + d_i  \,, \nn
\ea
with $\sum q_I^i d_i = c_I$ and such that ${\bf q_I} \perp {\bf v}$. For the submanifold $F$ to be Lagrangian with regard to the K\"ahler form $\omega = \sum_i d|X_i|^2 \wedge d\theta^i$, the angle valued 4-tuple ${\bf \theta}$ restricted to $F$ must satisfy ${\bf \theta} \perp {\bf v}$, and consequently, by counting dimensions, ${\bf \theta}|_F$ can be expanded in ${\bf q_I}$, ${\bf \theta}|_{F} = \phi^I {\bf q_I}$. For the submanifold to be special Lagrangian, $\sum_i q_I^i = 0$ \cite{AV}. Among these submanifolds, the Lagrangians whose base ends on the toric skeleton play a distinguished role: they have topology $\IR^2 \times S^1$, coordinatized by $s\in \IR$ and the angles $\phi_1$ and $\phi_2$, and cannot move off the skeleton without developing a boundary. 
Let us consider as an example the special Lagrangian defined by the vectors ${\bf q_1} = (0,1,-1,0)$ and ${\bf q_2} = (1,0,-1,0)$ with $c_1 = c>t$ and $c_2 = 0$. The corresponding line in the base is given by 
\ba
|X_2|^2 - |X_3|^2 &=& c  \nn\,, \\
|X_1|^2 - |X_3|^2 &=& 0  \nn  \,,
\ea
and it is easy to see that it ends on the edge of the toric skeleton extending along the $|X_2|^2$ axis. The torus fibered over this line is coordinatized by two angles $\phi_1$ and $\phi_2$ such that 
\ba
\theta_1 = \phi_2  \,, \hspace{0.3cm}\theta_2= \phi_1 \,,  \hspace{0.3cm} \theta_3 = - \phi_1 - \phi_2 \,,  \hspace{0.3cm} \theta_4 =0  \,.\nn
\ea
Note that the $S^1$ coordinatized by $\phi_1$ degenerates on the edge, whereas the $S^1$ coordinatized by $\phi_2$ is not contractible.

We review these well-known facts here to make the following observations, which will be important in the following:
\begin{itemize}
\item{The distinguished Lagrangians can be chosen to have one of the $c_I$ vanishing, say $c_2$. We can therefore add multiples of ${\bf q_2}$ to ${\bf q_1}$ to describe the same Lagrangian, but with $\phi_1$ and $\phi_2$ now coordinatizing different cycles in the $T^2$ fiber.}
\item{The non-contractible $S^1$ can be coordinatized by the phase of those two variables $X_i$ for which $|X_i|^2 \rightarrow \infty$ along the edge on which the brane ends.}
\item{By invoking $\sum_i Q_i |X_i|^2 =t$, we find to each ${\bf q_1}$ an equivalent ${\bf {\tilde q_1}}$ with $c_1$ shifted by $\pm t$.}
\end{itemize}

The moduli space of branes on the distinguished special Lagrangian submanifolds is a cylinder: it is given by the position along an edge, encoded in the area $e^{-r}$ of a holomorphic disc ending on the brane, together with a phase given by the Wilson loop around the $S^1$, specifying the flat bundle over the Lagrangian,
\ba
t = e^{-r}\, \tr P \exp \oint A \,,\nn
\ea
where $\tr$ specifies the trace in the fundamental representation (we will later use the notation $\Tr_{\!\!R}$ for a trace in the representation $R$).
A crucial observation for all of the following is that at a vertex, the entire fiber shrinks to zero, and we hence cannot naively follow the brane through the vertex from one edge to another. From the A-model point of view, the complete moduli space hence appears disconnected, consisting of a copy of a cylinder per edge \cite{hori}. In \cite{AV,AKV}, branes ending on different edges are referred to as different phases of the theory. One main purpose of this paper is to study how the open topological string partition functions corresponding to these different phases are related, following proposals of \cite{ADKMV} and \cite{ANV}. The title of the paper gives away the punch line.

To completely specify the open A-model on non-compact geometries, an additional integer must be specified for each non-compact brane in the geometry. This integer was referred to as a framing choice in \cite{AKV}, as this is what it corresponds to in the target space description of the open topological string given by Chern-Simons theory.

\subsection{Winding vs. representation basis in the A-model} \label{s:winding}
The free energy of the open topological A-model for a single stack of branes has the expansion 
\ba
F_{open}=F_{k_1 k_2\ldots} \,t_1^{k_1} \cdots t_n^{k_n} 
\cdots \,,  \nn
\ea
where the expansion is in complexified Wilson loops,
$V = P\exp \oint A$ and $t_n = e^{-nr} \tr V^n$. $k_n$ indicates the number of boundaries ending on the brane with winding number $n$. In the following, we will use the abbreviated notation
\ba
t^{{\bf k}} = \prod_{i=1}^{\infty} (t_i)^{k_i} \,. \nn
\ea
The partition function $Z_{open} = \exp F_{open}$ can also be expanded in Wilson loops, but note that the coefficients $Z_{{\bf k}}$ receive contributions from various $F_{{\bf k}}$. We refer to $Z_{{\bf k}} $ as the expansion coefficients of $Z$ in the winding basis. Due to the Frobenius relation
\ba  \label{Frobenius}
\prod_{n} (\tr V^n)^{k_n} = \sum_R \chi_R(C_{{\bf k}}) \Tr_{\! \! R} V
\ea
in the representation theory of the symmetric group (see appendix \ref{assorted} for some assorted comments on this formula), where $C_{{\bf k}}$ specifies the conjugacy class of elements with $k_i$ $i$-cycles, an expansion in a different basis, called the representation basis, is also possible. The representation basis arises naturally in the vertex formalism. The two representations are related via
\ba \label{windingandrep}
\sum_{{\bf k}} c_{\bf k} Z_{{\bf k}} \, t^{\bf k} = \sum_R Z_R \,e^{-|R|r} \,\Tr_{\!\! R} V \,,
\ea
(if we take the sum over ${\bf k}$ to include ${\bf k}=0$, corresponding to the closed string amplitude, then we must include the trivial representation on the RHS), where we have defined
\ba
Z_R = \sum_{{\bf k}} c_{\bf k}Z_{{\bf k}}\, \chi_R(C_{{\bf k}}) \,. \label{defzr}
\ea
We have introduced the formal product $c_{\bf k}=\prod_i (i^{k_i} k_i!)^{-1}$ in the definition of $Z_{\bf k}$ for later convenience, such that 
\ba 
c_{{\bf k}} &=& \frac{|C_{{\bf k}}|}{|G|} \,, \nn
\ea 
where $|C_{{\bf k}}|$ denotes the number of elements in the conjugacy class $C_{{\bf k}}$; in $S_n$, this is given by
\ba
|C_{{\bf k}}| = \frac{n!}{1^{k_1} k_1! 2^{k_2} k_2! \cdots n^{k_n} k_n!} \,.\nn
\ea
Note the following: the sum on the LHS of (\ref{windingandrep}) is over partitions ${\bf k}$ of $d$, representing elements of the symmetric group $S_d$, for some $d$. For $d>n$, with $n$ the rank of $V$, the polynomials $\prod_n (\tr V^n)^{k_n}$ in the eigenvalues of $V$ are not linearly independent.  To allow the unambiguous definition of the coefficients $Z_{\bf k}$ for arbitrarily large $d$, we must hence consider the limit $n \rightarrow \infty$. On the RHS, this is reflected in the fact that the contribution for representations $R$ whose Young tableaux has more rows than the rank $n$ of $V$ vanishes. For $d>n$, the sum hence does not receive contributions from all representations of $S_d$. 
As the $\Tr_{\!\! R}  V$ that don't vanish are linearly independent, the $Z_R$ are well-defined even at finite $n$.

We can invert (\ref{defzr}), and generally go back and forth between the winding and the representation basis, by invoking the two orthogonality relations for finite groups
\ba
\sum_{\bf k} c_{{\bf k}} \chi^*_{R_1}(C_{{\bf k}}) \chi_{R_2}(C_{{\bf k}}) &=& \delta_{R_1 R_2} \,, \nn \\
\sum_R \chi^*_R(C_{\bf{k_1}}) \chi_R(C_{\bf{k_2}}) &=& \frac{\delta_{{\bf k_1} {\bf k_2}}}{c_{\bf{k_1}}} \,. \nn
\ea
With the choice of normalization in (\ref{defzr}), we obtain the inverse relation
\ba
Z_{{\bf k}} = \sum_R Z_R \,\chi_R(C_{{\bf k}}) \,. \label{zkfromzr}
\ea

A useful fact from representation theory is that a polynomial $\chi_R(\bx)$, the character polynomial for the representation $R$ of the symmetric group, exists such that $\chi_R (C_\bk) = \chi_R(\bx)|_{\bx = \bk}$. Rewriting (\ref{zkfromzr}) as
\ba \label{towardsbf}
Z({\bf x})|_{\bx=\bk} = \sum_R Z_R \,\chi_R(\bx)|_{\bx=\bk} \,, 
\ea
we recognize that the LHS of (\ref{zkfromzr}) naturally lives in a bosonic Hilbert space, and the RHS in the isomorphic 0 charge fermionic Hilbert space, in the following sense \cite{AKMV}. The Hilbert space spanned by the modes $\alpha_n$ of a boson, satisfying the Heisenberg algebra $[\alpha_m,\alpha_n]=\delta_{m+n,0}$, and such that $\alpha_n |0\rangle = 0$ for $n>0$,  is isomorphic to the polynomial ring $\IC[x_1, \ldots]$, with $\alpha_{-n}$ for $n>0$ acting as multiplication by $x_n$ and $\alpha_n$ as differentiation. Hence, we can introduce the state
\ba
Z({\bf x}) &=& Z(\alpha_{-1},\ldots) |0\rangle \,, \nn
\ea
and the evaluation map is effected by considering the matrix element with the coherent state $| \bk \rangle = e^{\sum k_n \alpha_{-n}} |0 \rangle$,
\ba
Z({\bf x}) |_{\bx = \bk} &=& \langle \bk |Z(\alpha_{-1},\ldots) |0\rangle \,. \nn
\ea
The bosonic Hilbert space is isomorphic to the 0 charge sector of the fermionic Hilbert space spanned by the modes of a fermion satisfying the Clifford algebra $\{\psi_m, \psi^*_n \} = \delta_{m+n,0}$, with $\psi_m | 0 \rangle = \psi^*_m |0\rangle =0$ for $m > 0$. A basis for this space can be introduced whose basis vectors are in one to one relation to representations $R$ of the symmetric group (see e.g. \cite{miwa}):
\ba
\lefteqn{|(m_1 , \ldots, m_r  | n_1, \ldots, n_r ) \rangle =} \nn\\
& & (-1)^{\sum_{i=1}^r n_i + r(r-1)/2}\psi_{-m_1 - \frac{1}{2}} \ldots \psi_{-m_r- \frac{1}{2}} \psi^*_{-n_1- \frac{1}{2}} \ldots \psi^*_{-n_r- \frac{1}{2}} | 0 \rangle  \,,\nn
\ea
with $m_1 > \ldots > m_r$, $n_1> \ldots > n_r$, where we have specified the representation in terms of its Young diagram, indicating the number of boxes (to the right $|$ below ) the diagonal. E.g., for the hook representation with $m+1$ horizontal and $n+1$ vertical boxes, the corresponding state is
\ba
|(m|n) \rangle = (-1)^{n} \psi_{-m-\frac{1}{2}} \psi^*_{-n-\frac{1}{2} }  | 0 \rangle \,, \nn
\ea
and
\ba
\langle (m|n) |= (-1)^{n} \langle 0 | \psi_{n+\frac{1}{2}} \psi^*_{m+\frac{1}{2} }   \,. \nn
\ea
Under the isomorphism between the bosonic and fermionic Hilbert space, these basis vectors are mapped to the respective character polynomials (these are known to form a basis of the ring of polynomials),
\ba
\langle 0 | e^{\sum x_n \alpha_{n}} | R \rangle = \chi_R (\bx) \,.  \nn
\ea
The relation (\ref{towardsbf}) can hence be rewritten as
\ba
Z(\alpha_{-1},\ldots) |0\rangle &=& \sum_R | R \rangle \langle R | Z \rangle \,. \nn
\ea
This relation proves very powerful for the case of $\IC^3$, as \cite{ADKMV} show that the state $|Z \rangle$ for this geometry has a very simple fermionic representation as the exponential of a sum of bilinears in fermionic modes. All the information of the vertex is hence captured in the coefficients of these bilinears. In the following subsection, we demonstrate that this continues to hold for the conifold. It would be interesting to demonstrate, using the gluing rules of the topological vertex, that $|Z \rangle$ has this form for any toric target space.

Finally, note that we can obtain the full partition function by shifting the creation operators in the definition of coherent states by the open string moduli $t_n = e^{-x_n}$,
\ba
|t\rangle &=& \sum_{\bk}  e^{\sum k_n ( \alpha_n - x_n)}| 0 \rangle \,, \nn
\ea
such that $Z(t) = \langle t | Z \rangle$. 
 
\subsection{The fermion bilinear ansatz for $|Z\rangle$} \label{s:bilinear}
\begin{figure}[h]
\psfrag{a}{$\alpha$}
\psfrag{b}{$\beta$}
\psfrag{c}{$\gamma$}
\begin{center}
\epsfig{file=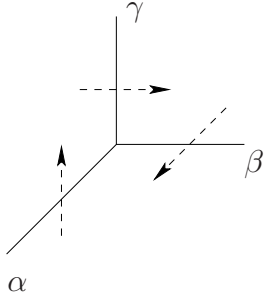, height=1.5in}
\end{center}
\caption{\small Vertex with standard framing. \label{fig:vertex}}
\end{figure}

The engine behind all the computations performed in this paper will be the topological vertex \cite{AKMV},
\begin{eqnarray}
C_{\gamma \beta \alpha} = q^{\frac{\kappa(\gamma)}{2}} \alpha \sum_\eta \frac{\gamma^t}{\eta}(\alpha) \frac{\beta}{\eta}(\alpha^t) \,. \label{vertex}
\end{eqnarray}
On the RHS, Greek letters are used to denote both representations of the symmetric group and the corresponding Schur functions.
See appendix \ref{apschur} for further notation as well as the many Schur function identities we will use in this subsection. The topological vertex is a function of the string coupling via $q=e^{g_s}$. As written, (\ref{vertex}) corresponds to the framing depicted in figure \ref{fig:vertex}. General framing comes with a factor of $(-1)^{\sum n_i |\alpha_i|} q^{\sum n_i \frac{\kappa(\alpha_i)}{2}}$ (see \cite{AKMV} for the definition of $n_i$).

\begin{figure}[h]
\psfrag{a}{$\alpha$}
\psfrag{b}{$\beta$}
\psfrag{c}{$\gamma$}
\psfrag{d}{$\delta$}
\begin{center}
\epsfig{file=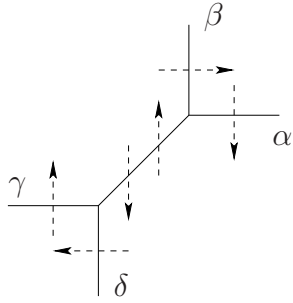, height=1.5in}
\end{center}
\caption{\small Conifold with non-standard framing. \label{fig:conifold}}
\end{figure}

The two geometries we will consider in this paper are $\IC^3$ and the conifold. The partition function $Z_{\gamma \beta \alpha}(\IC^3)$ is of course simply given by the vertex (\ref{vertex}). For the conifold, many of the formulae in the following sections simplify if we choose the framing indicated in figure \ref{fig:conifold}, i.e. $n=-1$ on the edges labeled by $\alpha$ and $\gamma$. 

Using the rules of the vertex, $Z_{\delta \gamma \beta \alpha}$ for the conifold is obtained as
\ba \label{typeiisum}
Z_{\delta \gamma \beta \alpha}  &=& (-1)^{|\gamma|} q^{-\frac{\kappa(\gamma)}{2}} C_{\rho \delta \gamma} \, (-1)^{|\alpha|} q^{-\frac{\kappa(\alpha)}{2}} C_{\rho^t \beta \alpha} (-1)^{|\rho|} Q^{|\rho |}\nn \\
  &=&  (-1)^{|\alpha|+|\gamma|}   \bigg[ \gamma^t \frac{\rho^t}{\eta_1}(\gamma) \frac{\delta}{\eta_1}(\gamma^t)  \bigg] (-1)^{|\rho|} Q^{|\rho |} \bigg[\alpha^t \frac{\rho}{\eta_2} (\alpha) \frac{\beta}{\eta_2} (\alpha^t) \bigg]\,, \nn
\ea
where we have introduced $Q=e^{-t}$. The sum over $\rho$ can be performed \cite{su2,sun,eguchi}, yielding
\ba
Z_{\delta \gamma \beta \alpha} &=& (-1)^{|\alpha|+|\gamma|}  \gamma^t \frac{\delta}{\eta_1}(\gamma^t)\sum_\alpha \frac{\rho^t}{\eta_1}(\gamma) \frac{\rho}{\eta_2}(\alpha) (-Q)^{|\rho|} \alpha^t \frac{\beta}{\eta_2} (\alpha^t) \nn\\ 
& =&  (-1)^{|\alpha|+|\gamma|} \gamma^t \frac{\delta}{\eta_1}(\gamma^t) \blb\gamma \alpha \brb_{Q} \alpha^t  \frac{\beta}{\eta_2} (\alpha^t)\sum_{\kappa} \frac{\eta_2^t}{\kappa^t} ({\gamma}) \frac{\eta_1^t}{\kappa}({\alpha}) (-Q)^{|\eta_1|+|\eta_2|-|\kappa|} \,. \nn
\ea
Again, for notation, see appendix \ref{apschur}.

With the state $|Z\rangle$ in the fermionic Hilbert space introduced above, these partition functions can be written as $Z_{\gamma \beta \delta}(\IC^3) =   \langle \gamma \beta \alpha| Z_{\IC^3} \rangle$ and $Z_{\delta \gamma \beta \alpha}({\rm conifold}) =  \langle \delta \gamma \beta \alpha| Z_{conifold} \rangle$. The authors of \cite{ADKMV} make the powerful conjecture that the state $ | Z \rangle$ can be written as an exponential in a sum of fermion bilinears (see also \cite{AKMV}),
\ba \label{vacuum}
|Z \rangle = \exp \left[ \sum_{i,j,n,m} a_{mn}^{ij}
\psi^i_{-m-\frac{1}{2}}
\psi^{j*}_{-n-\frac{1}{2}} \right]|0 \rangle  \,,
\ea
where independent modes $\psi^i, \psi^{i*}$ are introduced for each edge.\footnote{We have here chosen the normalization $\langle 0|Z\rangle =1$, i.e. not included the purely closed contribution. With this normalization, $Z_{\delta \gamma \beta \alpha} = \blb \cdot \cdot \brb \langle \delta \gamma \beta \alpha | Z \rangle$ for the conifold.}
If this is the case, then $|Z\rangle$ is determined by its matrix elements with states represented by hook tableaux \cite{ADKMV}, up to a few constants (see below). Recall that these states are given by
\ba
|(m|n) \rangle = (-1)^{n} \psi_{-m-\frac{1}{2}} \psi^*_{-n-\frac{1}{2} }  | 0 \rangle \,. \nn
\ea
By using the anticommutation relations of the fermionic modes, we obtain \cite{ADKMV}
\ba \label{oneinsert}
\langle (m|n) |Z \rangle = (-1)^{n} a^{ii}_{mn}  \,,
\ea
where the trivial representation is inserted at all but the $i$-th edge, and
\ba
\langle (p|p')^i (q|q')^j | Z \rangle = (-1)^{p' + q'} (
a^{ii}_{pp'} a^{jj}_{qq'} -a^{ij}_{pq'} a^{ji}_{qp'} ) \,, \label{2nontriv}
\ea
with non-trivial representations inserted at the two edges $i$ and $j$. Computing the LHS of (\ref{oneinsert}) via the topological vertex hence allows us to determine $a^{ii}$. The factorization 
\ba
\langle (p|p')^i (q|q')^j | Z \rangle - (-1)^{p' + q'}  a^{ii}_{pp'} a^{jj}_{qq'} = -(-1)^{p' + q'} a^{ij}_{pq'} a^{ji}_{qp'} \label{factorize}
\ea
is then a non-trivial check on the bilinear ansatz (\ref{vacuum}), and allows us to determine $a^{ij}$ up to constants (i.e. factors independent of the lower indices) which cancel in the product $a^{ij}_{mn} a^{ji}_{rs}$. To constrain these, we need to consider the matrix element of $| Z \rangle$ with three non-trivial representations, given by \cite{ADKMV}
\ba
\langle (m|m')^i (n|n')^j (r|r')^k | Z \rangle &=& (-1)^{m' + n' + r'} (a_{m m'}^{ii} a_{n n'}^{jj} a_{r r'}^{kk} + a_{mn'}^{ij} a_{n r'}^{jk} a_{r m'}^{ki} + a_{m r'}^{ik} a_{n m'}^{j i} a_{r n'}^{k j}\nn \\
& & - a_{m m'}^{ii} a_{n r'}^{jk} a_{r n'}^{kj} - a_{m r'}^{i k} a_{n n'}^{jj} a_{r m'}^{ki} - a_{m n'}^{ij} a_{nm'}^{ji} a_{r r'}^{kk} ) \,. \label{threenontrivial}
\ea
It will turn out that studying matrix elements of hook representations alone does not give rise to enough constraints to determine all such constants. Finding matrix elements that allow pinpointing the remaining few is an exercise we leave to the reader. 

In this section, after reviewing the computation of these coefficients, performed for $\IC^3$ in \cite{ADKMV}, we show that the conifold passes the factorization test as well, and we determine the coefficients $a_{ij}$ for this case. These coefficient will enter centrally in our study of the open string partition function on the conifold in upcoming sections. 

\paragraph{\mbox{\boldmath$\IC^3$}:} To obtain the coefficients $a^{ij}$ for $\IC^3$ \cite{ADKMV} we must specialize the expression (\ref{vertex}) for $\langle \gamma \beta \alpha | Z_{\IC^3} \rangle$ to the case $\alpha = (\alpha_1| \alpha_2)$, and $\beta = \gamma = \cdot$, or $\beta = (\beta_1|\beta_2)$, $\gamma=\cdot$, respectively. 

With merely one representation non-trivial, (\ref{vertex}) reduces to the Schur function of that representation,
\ba
\langle \alpha | Z_{\IC^3} \rangle = \alpha  \,. \nn
\ea
By the cyclic symmetry of the vertex, this result is independent of which of the three edges of the toric skeleton of $\IC^3$ the brane is ending on. Specializing to hook representations $\alpha = (\alpha_1|\alpha_2)$, we hence obtain
\ba \label{vertex11}
a_{\alpha_1 \alpha_2}^{ii} &=& (-1)^{\alpha_2} (\alpha_1|\alpha_2) \,.
\ea
As we demonstrate in (\ref{schur}) of the appendix, the Schur function for a hook representation evaluates to
\ba
(\alpha_1|\alpha_2)&=&  \frac{q^{\frac{\alpha_1(\alpha_1+1)-\alpha_2(\alpha_2+1)}{4}}}{[\alpha_1]![\alpha_2]![\alpha_1+\alpha_2+1]} \,. \nn
\ea

For two representations non-trivial,
\ba
\langle \beta \alpha \cdot| Z_{\IC^3} \rangle  &=& q^{\frac{\kappa(\beta)}{2}}\sum_\eta \frac{\beta^t}{\eta} \frac{\alpha}{\eta} \,. \nn
\ea
To demonstrate that this expression factorizes, as required by (\ref{factorize}), we use the following relation for skew Schur polynomials, derived in appendix \ref{apschur},
\ba
\frac{(\alpha_1|\alpha_2)}{(\eta_1|\eta_2)} &=& (\alpha_1-\eta_1-1|0)(0|\alpha_2-\eta_2-1)  \,. \label{skewschursimp}
\ea
We now easily obtain
\ba
\langle \beta \alpha \cdot| Z_{\IC^3} \rangle  &=&\beta \, \alpha + \sum_{\eta_1=0}^{\min(\beta_2,\alpha_1)} (\beta_2 - \eta_1-1|0) (\alpha_1 - \eta_1-1|0) q^{-\frac{1}{2}\hkappa(\beta_2)} \times \nn \\
&& \hspace{2cm} \times \sum_{\eta_2=0}^{\min(\beta_1,\alpha_2)} (0| \beta_1-\eta_2-1)(0|\alpha_2 - \eta_2-1) q^{\frac{1}{2} \hkappa(\beta_1)}  \nn \,.
\ea
where we have defined
\ba
\kappa(\alpha_1|\alpha_2) &=& \alpha_1 (\alpha_1+1) - \alpha_2 (\alpha_2 +1) \nn\\
&=& \hkappa(\alpha_1) - \hkappa(\alpha_2) \,.\nn
\ea
This yields the following expression for $a^{ij}$ and $a^{ji}$,
\ba (-1)^{\beta_2} a^{ij}_{\alpha_1 \beta_2} &=& C^{ij} \sum_{\eta_1=0}^{\min(\beta_2,\alpha_1)} (\beta_2 - \eta_1-1|0) (\alpha_1 - \eta_1-1|0) q^{-\frac{1}{2}\hkappa(\beta_2)}  \,,  \label{vertex12}\\
(-1)^{\alpha_2} a^{ji}_{\beta_1 \alpha_2} &=& C^{ji} \sum_{\eta_2=0}^{\min(\beta_1,\alpha_2)} (0| \beta_1-\eta_2-1)(0|\alpha_2 - \eta_2-1) q^{\frac{1}{2} \hkappa(\beta_1)}  \,, \nn
\ea
with $C^{ij} C^{ji} =-1$, where the ordered pair $(i,j) \in \{(1,2), (2,3), (3,1) \}$. By invoking (\ref{threenontrivial}) for low representations, we find
\ba
(C^{ij})^6 = q   \,. \nn
\ea

\paragraph{The conifold:} The conifold partition function with a brane ending on one edge is given by setting all but one of the representations in $\langle \delta \gamma \beta \alpha | Z_{conifold} \rangle$ to the trivial representation. The four choices lead to two types of expressions,
\ba
Z_{\cdot \cdot \cdot \alpha} &=& (-1)^{|\alpha|} \alpha^t \blb \alpha \cdot\brb \nn\\
Z_{\cdot \cdot \alpha \cdot} &=& \frac{\alpha}{\eta} \eta^t (-Q)^{|\eta|} \blb\cdot \cdot \brb \, = \, \alpha \blb\alpha^t \cdot\brb \nn\\
Z_{\cdot \alpha  \cdot \cdot} &=& (-1)^{|\alpha|}\alpha^t \blb\alpha \cdot\brb  \nn \\
Z_{\alpha \cdot \cdot \cdot} &=& \frac{\alpha}{\eta} \eta^t (-Q)^{|\eta|} \blb\cdot \cdot \brb \, = \, \alpha \blb \alpha^t \cdot\brb  \nn
\ea
reflecting the symmetry of (\ref{momentcfld}) under the joint transposition
\ba 
|X_1|^2 \leftrightarrow |X_2|^2 &,& |X_3|^2 \leftrightarrow |X_4|^2 \,.\nn
\ea
Labeling the 4 patches $Z_{\delta \gamma \beta \alpha}$ from 1 to 4, $Z_{4321}$, this corresponds to the symmetry
\ba \label{symmetry}
1 \leftrightarrow 3 &,& 2 \leftrightarrow 4 \,.
\ea
By specializing to hook representations, we can read off the coefficients of the fermion bilinears in accordance with (\ref{oneinsert}) as
\ba
a_{\alpha_1 \alpha_2}^{11} = a_{\alpha_1 \alpha_2}^{33} &=& (-1)^{\alpha_1+1} (\alpha_2 |\alpha_1) \{(\alpha_1 |\alpha_2) \cdot\}  \label{a11} 
\ea
and
\ba
a_{\alpha_1 \alpha_2}^{22} = a_{\alpha_1 \alpha_2}^{44} &=& (-1)^{\alpha_2} (\alpha_1|\alpha_2) \{(\alpha_2|\alpha_1) \cdot \}  \label{answer}    \,.
\ea
The RHS of the above expressions is easy to evaluate. We have already evaluated $(\alpha_1|\alpha_2)$ above. $\{ \alpha \cdot \}$ for hook representations is determined in appendix \ref{apschur}, see eq. (\ref{bracketone}),
\ba
\{ (\alpha_1|\alpha_2)  \cdot \} &=&  \prod_{i=-\alpha_2}^{\alpha_1} (1-Qq^i)\,. \nn
\ea

For two non-trivial representations, keeping the symmetry (\ref{symmetry}) in mind, 4 inequivalent configurations arise which constitute two pairs related by flop transitions.
\begin{figure}[h]
\psfrag{a}{$\alpha$}
\psfrag{b}{$\beta$}
\psfrag{1}{$1.$}
\psfrag{2}{$2.$}
\psfrag{3}{$3.$}
\psfrag{4}{$4.$}
\begin{center}
\epsfig{file=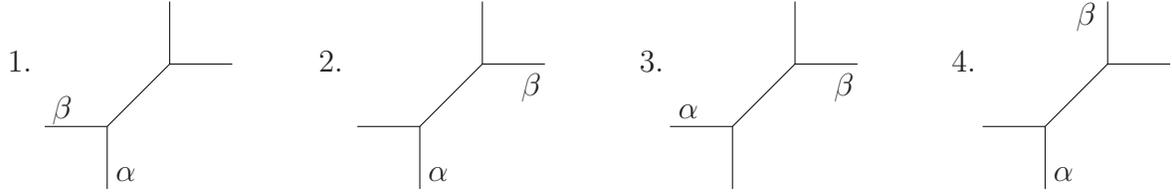, height=1in}
\end{center}
\caption{\small Four inequivalent configurations with two branes. \label{fig:fourtwos}}
\end{figure}

For configuration 1, we obtain
\ba
Z_{\alpha \beta  \cdot \cdot} / \blb \cdot \cdot \brb & =&(-1)^{|\beta|}\beta^t  \{\beta \cdot \}_{Q}   \sum_{\eta} \frac{\alpha}{\eta}(\beta^t) \eta^t (-Q)^{|\eta|} \nn \\
&=& \alpha \{\alpha^t \cdot\}_{Q} (-1)^{|\beta|} \beta^t \{\beta \cdot \}_{Q}+  (-1)^{|\beta|} q^{-\frac{\kappa(\beta)}{2}}\{\beta \cdot \}_{Q}  \sum_{\nu} \frac{\alpha^t}{\nu^t} (-Q)^{|\alpha|-|\nu|} \, q^{\frac{\kappa(\nu)}{2}} \sum_{\kappa \neq \cdot} \frac{\nu^t}{\kappa} \frac{\beta}{\kappa} \nn \,.
\ea
The steps between the first and second line, which are representative of the types of manipulations which enter in this subsection, are carefully spelled out in (\ref{2alongsame}) of the appendix. To factorize the second term as required by (\ref{factorize}), we specialize it to hook representations, $\alpha=(\alpha_1| \alpha_2)$, $\beta=(\beta_1| \beta_2)$, and use the identity
\ba\frac{(\alpha_1|\alpha_2)}{(\eta_1|\eta_2)} &=&  (\alpha_1-\eta_1-1|0) (0|\alpha_2-\eta_2-1) \,
\ea
derived in appendix \ref{apschur}.
We then easily find
\ba
(-1)^{\beta_2}a^{43}_{\alpha_1 \beta_2} &=& C^{43} (-1)^{\beta_2} q^{\frac{1}{2}\hkappa(\beta_2)}   \{(0|\beta_2) \cdot \} \sum_{\nu_1=0}^{\alpha_1} (0|\alpha_1 - \nu_1-1)  (-Q)^{\alpha_1 - \nu_1} \times \nn\\
& & \hspace{4.3cm} \times q^{\frac{1}{2}\hkappa(\nu_1)} \sum_{\kappa_2=0}^{\min(\beta_2, \nu_2)} (0|\nu_1 - \kappa_2 -1 ) (0| \beta_2 - \kappa_2 -1) \nn \\
(-1)^{\alpha_2} a^{34}_{\beta_1 \alpha_2} &=& C^{34}  (-1)^{\beta_1} q^{-\frac{1}{2}\hkappa(\beta_1)}  \{(\beta_1|0) \cdot \}  \sum_{\nu_2=0}^{\alpha_2} 
(\alpha_2- \nu_2-1|0) (-Q)^{\alpha_2 - \nu_2} \times \nn \\
& & \hspace{4.3cm} \times q^{-\frac{1}{2}\hkappa(\nu_2)} \sum_{\kappa_1=0}^{\min(\beta_1, \nu_1)} (\nu_2 - \kappa_1 -1|0 ) ( \beta_1 - \kappa_1 -1|0) \nn \,,
\ea
with $C^{43}  C^{34} = \frac{1}{(1-Q)}$.

For configuration 2, 
\ba
Z_{\alpha \cdot  \cdot \beta} / \blb \cdot \cdot \brb &=& (-1)^{|\beta|} \beta^t \{ \beta \cdot \} \sum_\eta \frac{\alpha}{\eta}  \eta^t(\beta) (-Q)^{|\eta|}  \,, \nn
\ea
and upon specializing to hook representations, a similar calculation as above yields
\ba
(-1)^{\beta_2} a_{\alpha_1 \beta_2}^{41} &=& C^{41} (-1)^{\beta_2}  \{(0|\beta_2) \cdot\} \sum_{\eta_1 =0}^{\alpha_1} (-Q)^{\eta_1} (\alpha_1 - \eta_1 -1|0) \times \nn \\
& & \hspace{4.3cm} \times q^{-\frac{1}{2}\hkappa(\eta_1)}\sum_{\kappa_1=0}^{\min(\beta_2,\eta_1)} (\beta_2-\kappa_1-1|0)(\eta_1-\kappa_1-1|0) \nn \\ 
(-1)^{\alpha_2} a_{\beta_1 \alpha_2}^{14} &=& C^{14} (-1)^{\beta_1} \{(\beta_1|0) \cdot\} \sum_{\eta_2 =0}^{\alpha_2} (-Q)^{\eta_2} (0| \alpha_2 - \eta_2 -1) \times \nn \\
& & \hspace{4.3cm}\times q^{\frac{1}{2}\hkappa(\eta_2)} \sum_{\kappa_2=0}^{\min(\beta_1,\eta_2)} (0|\beta_1-\kappa_2-1)(0|\eta_2-\kappa_2-1) \,, \nn 
\ea
with $C^{14} C^{41} = -\frac{Q}{(1-Q)}$.

Configuration 3 gives rise to
\ba
(-1)^{|\alpha| + |\beta|} Z_{\cdot \alpha  \cdot \beta}/ \blb \cdot \cdot \brb &=& \alpha^t \{ \alpha \beta \} \beta^t   \,.\nn
\ea
Specializing to hook representations leaves us with
\ba
(-1)^{|\alpha| + |\beta|}Z_{\cdot \alpha | \cdot \beta} / \blb \cdot \cdot \brb&=& \alpha^t \beta^t \{(\alpha_1|\alpha_2)(\beta_1|\beta_2)\}_Q \nn \\
&=& \alpha^t \beta^t  \{ \alpha \cdot \}_Q \{\beta \cdot\}_Q \frac{(1-Qq^{\alpha_1+\beta_1+1})(1-Qq^{-(\alpha_2+\beta_2+1)})}{(1-Qq^{\alpha_1-\beta_2})(1-Qq^{\beta_1-\alpha_2})} \,. \nonumber
\ea
A gratifying cancellation, see (\ref{disent}), now disentangles $\alpha_1, \alpha_2$ in $\alpha$, and likewise for $\beta$, resulting in
\ba
a^{31}_{\alpha_1 \beta_2}&=& C^{31} (-1)^{\alpha_1}(\beta_2|\alpha_1) \{(\alpha_1|\beta_2) \cdot\} \frac{q^{\alpha_1+\frac{1}{2}}-q^{-\beta_2-\frac{1}{2}}}{1-Q q^{\alpha_1-\beta_2}} \,,\nn\\
a^{13}_{\beta_1 \alpha_2}&=& C^{13}(-1)^{ \beta_1} (\alpha_2|\beta_1) \{(\beta_1|\alpha_2) \cdot\} \frac{q^{\beta_1+\frac{1}{2}}-q^{-\alpha_2-\frac{1}{2}}}{1-Q q^{\beta_1-\alpha_2}} \,,\nn
\ea
with $C^{13} C^{31} = Q$. By the symmetries of the conifold, we know that $C^{13} = C^{31}$, hence $C^{13} = c_{13} \sqrt{Q}$, with $c_{13}^2=1$.

Finally, for configuration 4, we get
\ba
Z_{\alpha \cdot \beta \cdot} / \blb \cdot \cdot \brb &=&  \frac{\alpha}{\eta}   \frac{\beta}{\nu} \sum_{\kappa} \frac{\nu^t}{\kappa^t}  \frac{\eta^t}{\kappa}(-Q)^{|\eta|+|\nu|-|\kappa|} \nn\\
&=&  \alpha \{ \alpha^t \cdot \}_Q \,\beta \{ \beta^t \cdot \}_Q  + \sum_{\eta, \nu, \kappa \neq \cdot} \frac{\alpha}{\eta} \frac{\eta^t}{\kappa} \frac{\beta}{\nu} \frac{\nu^t}{\kappa^t} (-Q)^{|\eta| + |\nu| - |\kappa|} \nn \,.
\ea
This yields
\ba
(-1)^{\beta_2} a^{42}_{\alpha_1 \beta_2}& =& C^{42} \sum_{\eta_1=0}^{\alpha_1} \sum_{\nu_2=0}^{\beta_2} \sum_{\kappa_2=0}^{\min(\eta_1,\nu_2)} (\alpha_1-\eta_1-1|0)(0|\eta_1-\kappa_2-1) \times \nn \\
& & \hspace{2.5 cm} \times (\nu_2-\kappa_2-1|0)(0|\beta_2-\nu_2-1) (-Q)^{\eta_1 + \nu_2-\kappa_2}  \,, \nn\\
(-1)^{\alpha_2} a^{24}_{\beta_1 \alpha_2} &=& C^{24}
 \sum_{\eta_2=0}^{\alpha_2} \sum_{\nu_1=0}^{\beta_1} \sum_{\kappa_1=0}^{\min(\eta_2,\nu_1)}(0|\alpha_2-\eta_2-1) (\eta_2-\kappa_1-1|0) \times \nn\\
 & & \hspace{2.5cm} \times(0|\nu_1-\kappa_1-1)(\beta_1-\nu_1-1|0) (-Q)^{\eta_2 + \nu_1-\kappa_1} \,, \nn
\ea
with $C^{24} C^{42} = Q$. By the symmetries of the conifold, $C^{24} = C^{42}$, allowing us to conclude $C^{24} = c^{24} \sqrt{Q}$ with $(c^{24})^2=1$.
 
By invoking (\ref{threenontrivial}) for low representations, we can determine the following relations between the constants $C^{ij}$,
\ba
C^{14} &=& - \sqrt{qQ} C^{34} \,,  \nn \\
c^{13} c^{24} &=& -1 \nn  \,.
\ea 
 
Note that the coefficients $a^{ij}$ related by flop transitions are very similar. The factors they differ by are interpreted geometrically in \cite{strip}, see also \cite{flop}.
 
\section{The fermion operator} \label{s:fermionops}
The target space description of the open A-model is given by an instanton modified Chern-Simons action \cite{cswitten}. In fact, this description underlies \cite{allgenus,amer} the development of the vertex formalism. The target space description of the B-model is given by holomorphic Chern-Simons theory. On the mirrors to toric geometries, to which we return in section \ref{s:wave}, the dynamical field of this description is a chiral scalar \cite{AV, DV}. The authors of \cite{AKMV} and \cite{ADKMV} propose interpreting the modes $\alpha_n$ introduced above as arising in the mode expansion of this scalar field at the infinities of the target space geometry. The fermionization of the chiral scalar,
\ba
\psi(x) = \sum_{m=-\infty}^{\infty} \psi_{m-\frac{1}{2}} e^{-m x} \,,\nn
\ea
is associated with solitonic excitations of the theory, i.e. branes, such that $x$ is a D-brane modulus: from the A-model point of view, $e^{-x} = e^{-r} \tr V$. In this section, we wish to study insertions of this operator from an A-model point of view.

\subsection{One- and two-point functions of the fermion operator} \label{s:oneandtwo}
An insertion of the fermion operator $\psi^*$ yields the following,
\ba
\Psi^*(x) &=& \langle 0 |\psi_{\frac{1}{2}} \psi^*(x)|Z\rangle \nn \\
&=&1+ \sum_{n=0}^{\infty} \Tr_{\!\!(n|0)}(e^{-x}) \langle (n|0) | Z \rangle  \nn\\
&=& \sum_R Z_R \Tr_{\!\!R}V  \,. \nn
\ea
with $V=e^{-x}$. In writing the last sum over all representations $R$ (including the trivial representation), we have used that $\Tr_{\!\!R} V =0$ if the Young tableaux corresponding to the representation $R$ has more than rank $V$ non-vanishing rows. $\Psi^*(x)$ hence yields the open topological string partition function with $U(1)$ Wilson loop factors $t_n= e^{-nx}$. For the operator $\psi$, we get
\ba
\Psi(x) &=& \langle 0 |\psi^*_{\frac{1}{2}} \psi(x)| Z \rangle \nn\\ &=& 1+  \sum_{n=0}^{\infty} (-1)^{n+1} \langle (0|n) | Z\rangle  \Tr_{(n|0)} V\nn \\
&=& \sum_R (-1)^{|R|} Z_R \Tr_{\!\!R^t} V \,, \nn
\ea
with $|R|$ the number of boxes in the Young tableaux corresponding to the representation $R$. Substituting (\ref{defzr}) and using the relation $\chi_R (C_{\bf k})= (-1)^{|R| + |{\bf k}|}\chi_{R^t} (C_{\bf k})$ between characters of conjugate partitions , with $|{\bf k}| = \sum k_i$, we obtain
\ba
\Psi(x) &=& \sum_{\bf k} (-1)^{|{\bf k}|} c_{\bf k} Z_{\bf k} t^{\bf k} \,,
\ea
(a similar calculation also appears in \cite{AKMV}).
$\Psi(x)$ hence yields the open topological string partition function with $U(1)$ Wilson loop factors $t_n= e^{-nx}$ and each hole  weighted by a minus sign. Vafa \cite{vafanti} identifies this property as distinguishing a topological antibrane from a brane.

We conclude that inclusion of the operator $\psi^*(x)$, $\psi(x)$ respectively generate the open topological string partition function with a single brane/antibrane of modulus $e^{-x}$.

Note that a crucial step in this identification was that all coefficients $\langle R | Z \rangle$ with $\langle R|$ requiring more than one fermionic mode to generate could be dropped, by $\Tr_{\!\! R} V=0$ for such $R$. What about multiple insertions? Since we have introduced independent fermionic modes $\psi^i$ for each edge $i$, insertions on separate edges simply yield
\ba
\langle 0 |\psi^i_{\frac{1}{2}} \psi^j_{\frac{1}{2}} \psi^{j*}(x) \psi^{i*}(y) | Z \rangle &=& \sum_{R_i, R_j} Z_{R_i,R_j} \Tr_{\!\!R_i}(e^y) \Tr_{\!\!R_j}(e^x) \label{2pointdiflegs} \,,
\ea
where the sum again includes $R_i = \cdot$ and $R_j = \cdot$. The story becomes more involved for multiple insertions on the same edge. For two $\psi^*$ insertions, we obtain
\ba
\langle 0| \psi_{\frac{1}{2}} \psi_{\frac{3}{2}} \psi^*(x) \psi^*(y) | Z\rangle &=& e^x \big[ \Psi^*(y) - \sum_n \langle (n|1) | Z \rangle e^{-x-(n+1)y} + \label{2ptint} \\
&& \hspace{4cm}+\sum_{m < n} \langle (n,m|1,0) | Z \rangle e^{-(m+2)x - (n+1)y} \big] \nn \\ 
& &- e^y \big[ \Psi^*(x) - \sum_m \langle (m|1) | Z \rangle e^{-(m+1)x-y} + \nn \\ 
&& \hspace{4cm} + \sum_{n < m} \langle (m,n|1,0) | Z \rangle e^{-(m+1)x - (n+2)y} \big] \nn \,.
\ea
We would have naively expected all coefficients $\langle R | Z \rangle$ to be weighted by the corresponding Schur polynomials. E.g. for $\langle (m,n|1,0) | Z \rangle$, the corresponding Schur polynomial is
\ba 
\Tr_{\!\!(m,n|1,0)}V &=& s_{(m,n|1,0)}(e^{-x},e^{-y}) \nn\\
&=& e^{-r(x+y)} ( e^{-sx} + e^{-(s-1)x - y } + \ldots + e^{-sy} ) \nn \\
&=& e^{-(m+1)x - (n+2)y} + \ldots \nn \,,
\ea
with $r=n+2, s=m-n-1$, (so e.g. $s_{\tableau{5 2}}(x_1,x_2) = x_1^2 x_2^2 (x_1^3 + \ldots)$). It is tempting to interpret the terms that appear in equation (\ref{2ptint}) as the leading contributions to the polynomials in an expansion in $e^{-x}$, $e^{-y}$ respectively. Such an expansion would make sense if we fix the position of one brane while taking the other to infinity along the edge. A better understanding of the A-model interpretation of such higher point functions is clearly desirable.

The only higher point function which will enter in the discussion in this paper, in section \ref{s:transop}, arises from the insertion of one $\psi$ and one $\psi^*$ operator on the same edge, for which we obtain
\ba \label{brane-antibrane}
\langle 0 | \psi(x) \psi^*(y) | Z \rangle &=& \sum_{m,n} (-1)^n \langle (m|n) | Z \rangle e^{-(m+1) y -(n+1)x} + \frac{1}{e^{y} - e^{x}} \,.
\ea

\subsection{Evaluating the one- and two-point functions} \label{s:eval12}
Having determined the coefficients $a_{mn}^{ij}$ in the fermion bilinear expansion (\ref{vacuum}) of the vacuum for $\IC^3$ and the conifold, we can easily evaluate the one and two point functions of the fermion operator for these geometries.
For the one point function, we obtain
\ba
\langle 0 | \psi^*_{\frac{1}{2}} \psi(x) |Z\rangle
&=& 1- \sum_{n=0}^\infty a_{0n} \,e^{-(n+1)x}   \,. \nn
\ea
Evaluating (\ref{answer}) for branes ending on the edges labeled by $2$ and $4$ of the conifold, 
\ba 
a_{0n}^{22}=a_{0n}^{44} &=& (-1)^{n} (0|n) \, \{(n|0) \cdot\} \nn\\
&=& (-1)^{n} \frac{\prod_{i=1}^{n+1} (1-Qq^{i-1})}{[n+1]!} q^{-\frac{1}{4} (n+1)n}  \,, \nn
\ea
yields
\ba
\Psi^{2}(x)=\Psi^4(x)&=& \sum_{r=0}^{\infty} (-1)^r \frac{\prod_{i=1}^{r} (1-Qq^{i-1})}{[r]!} q^{-\frac{1}{4} (r-1)r} e^{-rx}  \label{sumrep}\\
&=& \sum_{r=0}^{\infty} \prod_{i=1}^{r} \frac{1-Qq^{i-1}}{1-q^i} q^{\frac{r}{2}} e^{-rx}  \nn \\
&=& \prod_{i=0}^{\infty} \frac{1-Qq^{i+\frac{1}{2}}e^{-x}}{1- q ^{i+\frac{1}{2}}e^{-x}}  \label{productrep24}\\
&=& \exp[ - \sum_{n=1}^{\infty} \frac{e^{-nx}}{n[n]}(1-Q^n)] \,, \nn
\ea
and by comparing (\ref{answer}) to (\ref{vertex11}), we see that we can obtain the one-point function for $\IC^3$ from this by setting $Q=0$, 
\ba
\Psi(x) =\exp[ - \sum_{n=1}^{\infty} \frac{e^{-nx}}{n[n]}] \,, \nn
\ea
thus reproducing the result of \cite{ADKMV}.

For branes ending on the edges $1$ and $3$, evaluating
\ba
a_{0n}^{11}=a_{0n}^{33} &=& - (n|0) \, \{(0|n) \cdot\} \nn
\ea
yields
\ba
\Psi^{1}(x)=\Psi^3(x)&=& \sum_{r=0}^{\infty}  \frac{\prod_{i=1}^{r} (1-Qq^{1-i})}{[r]!} q^{+\frac{1}{4} (r-1)r} e^{-rx} \label{oneptexpstar} \\
 &=& \sum_{r=0}^{\infty} \prod_{i=1}^{r} \frac{Q-q^{i-1}}{1-q^i} q^{\frac{r}{2}} e^{-rx} \nn \\
 &=& \prod_{i=0}^{\infty} \frac{1- q ^{i+\frac{1}{2}}e^{-x}}{1-Qq^{i+\frac{1}{2}}e^{-x}} \label{productrep13}\\
 &=& \exp[ \sum_{n=1}^{\infty} \frac{e^{-nx}}{n[n]}(1-Q^n)] \,. \nn
\ea
Analogously, we obtain
\ba
\langle 0 |\psi_{-\frac{1}{2}} \psi^*(x) |Z\rangle &=&  1 +  \sum_{n=0}^{\infty} a_{n0} \,e^{-(n+1)x} \nn
\ea
and with
\ba
a_{n0}^{22}= a_{n0}^{44}&=& - a_{0n}^{11} = -a_{0n}^{33} \,, \nn
\ea
we can evaluate the conjugate one point function to be
\ba
\Psi^{2*}(x)=\Psi^{4*}(x)&=& \Psi^{1}(x) = \Psi^{3}(x) \nn
\ea 
while
\ba
a_{n0}^{11}= a_{n0}^{33}&=& -a_{n0}^{22}= -a_{n0}^{44}\nn
\ea
and
\ba
\Psi^{1*}(x)=\Psi^{3*}(x)&=& \Psi^{2}(x) = \Psi^{4}(x) \,. \nn
\ea
The comparison of (\ref{answer}) to (\ref{vertex11}) again let's us determine the one point function $\Psi^*$ of $\psi^*$ for $\IC^3$ by setting $Q=0$ in $\Psi^{2,4*}$,
\ba
\Psi^*(x) &=& \exp[ \sum_{n=1}^{\infty} \frac{e^{-nx}}{n[n]}] \,\,=\,\, \frac{1}{\Psi(x)} \,.
\ea
Non-canonical framing shifts the power of $q$ in the sum representation (\ref{sumrep}) of the one point function, such that e.g. $\Psi^{1}$ for general framing $k$ (relative to the framing depicted in figure \ref{fig:conifold}) is given by
\ba
\Psi^{1(k)}(x) &=& \sum_{r=0}^{\infty}  \frac{\prod_{i=1}^{r} (1-Qq^{1-i})}{[r]!} q^{+\frac{1}{4} (r-1)r(1+2k)} (-1)^{rk} e^{-rx} \,. \nn
\ea
The representation of the one-point function as an infinite product, as in (\ref{productrep24}) and (\ref{productrep13}), appears to only be possible for the specific choices of framing we have been considering, depicted in figures \ref{fig:vertex} and \ref{fig:conifold}.

Our main focus in the next section will be the transformation properties of these one point functions. The authors of \cite{ADKMV} propose assigning the transformation properties to the fermion operators themselves. To study this proposal, we will, following \cite{ADKMV}, consider the transformation of the brane-antibrane correlator (\ref{brane-antibrane}). We choose this specific two point function for simplicity: for branes/antibranes ending on the same edge, it satisfies the following identity,
\ba \label{idbraneantibrane}
\langle 0 | \psi(y) \psi^*(x) | Z \rangle &=& \frac{1}{e^{y} - e^{x}} \Psi(y) \Psi^*(x) \,,
\ea
which simplifies its manipulation. \cite{ADKMV} showed that this identity holds for $\IC^3$. In appendix \ref{apschur}, we show that it holds for the conifold as well.

For brane and antibrane ending on separate edges, we easily find
\ba
\langle 0 | \psi^j(y) \psi^{i*}(x) | Z \rangle &=& \sum_{m,n=0}^{\infty} a_{mn}^{ij} e^{-(m+1)x} e^{-(n+1)y} \,. \nn
\ea

\section{The partition function as wave function} \label{s:wave}
\subsection{Brane placement as choice of polarization} \label{s:bpcp}
In \cite{cswitten}, Witten provides a target space description of the open topological string in the A-model as a Chern-Simons theory, modified by instanton corrections, living on the Lagrangian submanifold  wrapped by the brane. Given a Lagrangian with boundary, or a non-compact Lagrangian, interpreted as having a boundary at infinity, the partition function of this target space theory behaves as a wave function on the phase space given by the field configurations restricted to the boundary. As the authors of \cite{ANV} point out, this provides the backdrop to interpreting the open topological string partition function as a wave function. The non-compact Lagrangians introduced in section \ref{setup} have the topology of a solid torus, with a boundary $T^2$ at infinity. The phase space is given by the holonomy of the Chern-Simons gauge field around the two cycles of this torus. In a Hamiltonian framework, only one of these two variables is fixed as a boundary condition when evaluating the path integral of the target space action on the Lagrangian submanifold. Evaluating the path integral hence yields a wave function as a function of this boundary condition. By standard arguments in quantum mechanics, different choices of cycle yield wave functions which are related by canonical transformations, see appendix \ref{QM}. 

\begin{figure}[h]
\begin{center}
\epsfig{file=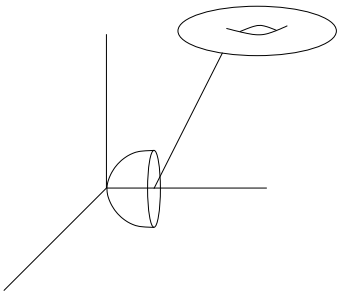, height=1.5in}
\end{center}
\caption{\small The boundary torus at infinity vs. the degenerate torus on which the worldsheet ends. \label{fig:t vs w}}
\end{figure}

From the point of view of the worldsheet, the open string partition function depends on an open string modulus, as explained in section \ref{setup}. It is natural to identify the dependence on this modulus with the boundary condition dependence in the target space picture. This modulus also requires a choice of cycle, however now at the center of the solid torus. The two relevant tori, at infinity for the target space description, and at the center of the solid torus and hence degenerate for the worldsheet, are sketched in figure \ref{fig:t vs w}. The only natural choice for the worldsheet theory is a non-contractible cycle of the solid torus. As we will explicitly see below, adding multiples of the contractible cycle to this choice corresponds to a change of framing from the point of view of the topological string (\cite{AKV}, see \cite{katz} for this integer choice appearing in a mathematical treatment of the open topological string; note that this integer choice has still not been identified from the worldsheet description of the open topological string). Choosing the contractible cycle as canonical variable on the other hand does not have a clear interpretation from the worldsheet point of view.

From the description of the moduli space of branes as a disconnected union of cylinders, as presented in section \ref{setup}, the story seems to end here, with a choice of polarization not accessible from the worldsheet description of the topological string. However, passing to the B-model via mirror symmetry, we obtain an improved understanding of the moduli space as a 4-punctured sphere, i.e. four {\it joined} cylinders \cite{AV}, see appendix \ref{s:tmg}. Given this modified picture of moduli space, we can determine the open topological string partition function for a brane close to a given puncture, corresponding to a Lagrangian ending on the corresponding edge from the A-model point of view, and consider following the partition function as we move along moduli space to a different puncture, i.e. to a different phase of the topological string from the A-model point of view. From the discussion above, we can thus obtain the open topological string partition function expressed in the unnatural coordinate corresponding to the complexified Wilson loop around a contractible cycle. To arrive at an expression for the partition function in natural topological string variables, we must hence perform a canonical transformation on the partition function. Thus, speaking laxly, different brane placements in the open topological string partition function are related via changes of polarization of phase space.

In the remaining parts of this section, we wish to substantiate this picture in the case of the conifold. We give a prescription for determining the canonical coordinates corresponding to the different punctures, and then verify that the open string partition functions for different brane placements transform into each other in accordance with this coordinate assignment.

\subsection{The assignment of coordinates} \label{s:assignment}
We introduce homogeneous variables $y_i = e^{-Y_i}$ on $\IC \IP^n$.
The Riemann surface representing the moduli space in the B-model \cite{HIV} is cut out of the ambient space $\IC\IP^n$ ($n=2$ for $\IC^3$, $n=3$ for the conifold) by linear equations in $Y_i$ (none for $\IC^3$, one for the conifold) and one homogeneous equation in $e^{-Y_i}$,
\ba
\sum e^{-Y_i} = 0 \,. \label{gconstraint}
\ea 
We impose the first set of equations (linear in $Y_i$) to obtain an extended phase space $M$, and, following \cite{ADKMV},  treat the equation homogeneous in $y_i$ as a constraint to be imposed after quantization.

For $\IC^3$, there is no linear equation in $Y_i$ to impose, and our proposal for the extended phase space $M$ descends from the space $\tilde{M}=\IC \IP^2$. As the $Y_i$ are the natural coordinates in the B-model \cite{HV}, the proper symplectic form on phase space is  
\ba
\omega &=& \pm \;d \log \frac{y_1}{y_3} \wedge d \log \frac{y_2}{y_3} \,. \label{symplectic form c3}
\ea
Note that $\omega$ is not defined when {\it any} of the coordinates $y_1, y_2, y_3$ vanish (in particular, $y_3$ is not distinguished in this definition, despite appearance). The extended phase space $M$ is hence properly identified with $\IC \IP^2$ with this locus removed, and is thus given by $\IC^* \times \IC^*$. We nevertheless refrain from introducing global coordinates on this space, as the relevant class of polarizations descend from foliations of patches of $\tilde{M}$, as described below.

For the conifold, the extended phase space $M$ descends from the quadric $\tilde{M}$
\ba
y_1 y_2 &=& y_3 y_4 e^{-t}
\ea
in $\IC \IP^3$. Consider the point $t=0$ in complex structure moduli space. By setting
\ba
y_1 = a b \;,\;\; y_2 = c d \;,\;\; y_3 = ad \;,\;\; y_4=bc \,,
\ea
we can identify this space with $\IC \IP^1 \times \IC \IP^1$, coordinatized as $(a:c) \times (b:d)$.\footnote{We thank Greg Moore for very helpful comments pertaining to the following discussion.} For arbitrary $t$, this discussion goes through after a rescaling of the $y_i$, hence yields $\IC \IP^1 \times \IC \IP^1$ with a shifted complex structure. For the symplectic considerations in this subsection, this shift is not relevant (note that the complex structure reenters in imposing the constraint equation on the Hilbert space). The proper symplectic form on phase space is  
\ba
\omega &=& \pm \;d \log \frac{a}{c} \wedge d \log \frac{b}{d} \,. \label{symplectic form conifold}
\ea
Again, $\omega$ is not defined when any of the coordinates $a, b, c, d$ vanish. We therefore identify $M$ with $\IC \IP^1 \times \IC\IP^1$ with this locus removed, again obtaining $\IC^* \times \IC^*$. 

An important comment is in order. Note that we are proposing a 2 complex dimensional phase space with a $(2,0)$ form as symplectic form, whereas our wave functions are to depend holomorphically on one complex variable. This is in marked contrast to the case of holomorphic polarization in geometric quantization, in which the symplectic form is a $(1,1)$ form, and a choice of polarization is a choice of complex structure. Instead, the correct approach in our study seems to involve treating the $Y_i$ as real variables in counting the dimension of phase space and choosing a symplectic form, but treating them as complex e.g. in performing the integrations involved in canonical transformations. 

The ambiguity can be traced back to the target space description of the open A-model as Chern-Simons theory. Conventional (real) Chern-Simons theory only captures the variations of the flat bundle on the brane. Complexifying the gauge field is to account for variations of the Lagrangian submanifold the brane is wrapping, but without doubling the degrees of freedom of the theory. We will see further indications that such a complexification is required in the following.

We first turn however to the relevant class of polarizations of the extended phase space. Recall that a polarization is a foliation of phase space by Lagrangian submanifolds. As announced above, we will consider Lagrangian foliations of $M$ that descend from foliations of patches of $\tilde{M}$. 

For $\IC^3$, we consider foliations of a given patch $y_i \neq 0$ of $\IC \IP^2$ by fixing one choice of homogeneous variable in that patch -- this corresponds to the exponential of the canonical coordinate -- while varying the other -- the exponential of the canonical momentum.

For the conifold, an additional subtlety arises. Concretely, in the patch $y_1 \neq 0$, the defining quadric equation for $\tilde{M}$ becomes
\ba
\frac{y_2}{y_1} &=& \frac{y_3}{y_1} \frac{y_4}{y_1} e^{-t} \,,
\ea
and we again consider foliations that arise by fixing one choice of homogeneous variable while varying another. We will find that the foliations with linear equations for the leaves, in the above example, this corresponds to fixing $\frac{y_3}{y_1}$ or $\frac{y_4}{y_1}$, give rise to the appropriate class of polarizations for our study.

Our task is now to assign a choice of polarization of phase space to each external edge of the toric skeleton of the A-model geometry. As discussed, these are in 1 to 1 correspondence with the punctures of the Riemann surface representing the moduli space of B-branes: the coordinates $y_i$ vanishing on the puncture map to the coordinates $|X_i|$ going to infinity along the corresponding leg. We identified this property in subsections \ref{setup} and \ref{s:bpcp} as the criterion for the associated phases to qualify as canonical coordinates (corresponding to a Wilson loop around a non-contractible $S^1$). It is hence natural to assign to each puncture a polarization that descends from a foliation of a patch of $\tilde{M}$ that contains the puncture. Following \cite{ADKMV}, we will impose the complementary constraint on the canonical momenta:
that they are chosen among the variables that correspond to a contractible $S^1$ on the leg of the toric diagram, hence do not vanish at the puncture.

These two criteria assign a subset of the polarizations introduced above to each puncture. We need to identify the sources of degeneracy within this subset, such that we can assign a unique choice of canonical coordinates to each of the partition functions determined in section \ref{s:eval12}. We make this identification with the help of the constraint equation (\ref{gconstraint}): each choice of polarization maps the LHS of this equation to a different differential operator. Each partition function calculated in section \ref{s:eval12} incorporates certain data (position of brane, brane vs. antibrane, etc.). By determining which partition function is annihilated by which operator, we can map this data to a choice of polarization.

\paragraph{\mbox{\boldmath$\IC^3$}:} Recall the equation of the Riemann surface describing the B-brane moduli space on the mirror of $\IC^3$,
\ba
e^{-Y_1} + e^{-Y_2}+e^{-Y_3} = 0 \label{3prs} \,.
\ea
At each puncture, $\re Y_i$ goes off to infinity for one $i$. At each puncture, we hence have the choice between polarizations descending from foliations of two patches of $\IC \IP^2$. At $P_1: (0:1:-1)$, e.g., we can consider the patch $y_2 \neq 0$. Then, following our prescription, the canonical coordinates corresponding to $P_1$ are $e^{-x} = \frac{y_1}{y_2}$, and $e^{-p} = \frac{y_3}{y_2}$. This choice is hence consistent with the positive sign in the definition (\ref{symplectic form c3}) of the symplectic form. The choice of coordinates corresponding to the patch $y_3 \neq 0$ corresponds to the negative sign. Choosing a sign, we can hence assign a polarization to each puncture, as presented in table \ref{preconcor}.
\begin{table}
\begin{tabular}{c|c||c|c||c|c}
\multicolumn{2}{c||}{$P_1: (0:1:-1)$} & \multicolumn{2}{c||}{$P_2:(1:0:-1)$} & \multicolumn{2}{c}{$P_3:(1:-1:0)$} \\ \hline
$p_1$ & $x_1$ & $p_2$ & $x_2$ & $p_3$ & $x_3$ \\ \hline
$-\log y_3/y_2$ & $-\log y_1/y_2$& $-\log y_1/ y_3 $ &$ - \log y_2 / y_3$ &  $ -\log y_2  /y_1$ & $-\log y_3 / y_1 $ \\ \hline
$-x$ & $p-x $ & $p$ & $x$ & $ x-p $ & $-p$  
\end{tabular} 
\caption{\small{Canonical variables for $\IC^3$. In the last row, $(p_1,x_1)$ and $(p_3,x_3)$ are expressed in terms of $(p_2,x_2)$. \label{preconcor}}}
\end{table}

Upon a choice of polarization, the LHS of the equation (\ref{3prs}), which we wish to interpret as a constraint equation on the Hilbert space, gets mapped to a differential operator \cite{ADKMV}. The three polarizations listed in table \ref{preconcor} yield the same constraint
\ba
1+ e^{-x} + e^{-p} &=& 0 \,, \nn
\ea
where $x$ is now interpreted as a multiplication operator, and $p \mapsto g_s \partial_x$. It is a consistency check for our choice of polarizations that the partition function,
\ba
\Psi(x) &=& \prod_{i=0}^{\infty} (1-q^{i+\frac{1}{2}}e^{-x})^{-1} \,,\nn
\ea
which coincides at all three punctures, i.e. for branes positioned at any of the three legs of the toric diagram of $\IC^3$, is annihilated by this operator, up to some factors of $q$ which are due to normal ordering ambiguities and which we must introduce by hand,
\ba
(1-q^{\frac{1}{2}}e^{-x} - e^{-g_s \partial_x}) \Psi(x) &=& 0 \,. \nn
\ea

Once we have assigned coordinates to the punctures as above, there are two operations on the coordinates which we can interpret physically with the help of the constraint equation.

If we consider coordinates consistent with the opposite sign of the symplectic form not by changing patches, but by inverting the sign of the momentum, $p \mapsto -p$, the constraint equation becomes
\ba
1+ e^{-x} + e^{p} &=& 0 \,, \nn
\ea
which annihilates
\ba
\Psi^*(x) &=& \prod_{i=0}^{\infty} (1-q^{i+\frac{1}{2}}e^{-x}) \,, \nn
\ea
again up to $q$-shifts,
\ba
(1-q^{-\frac{1}{2}}e^{-x} - e^{g_s \partial_x}) \Psi^*(x) &=& 0 \,. \nn
\ea
This choice hence distinguishes between $\Psi^*$ and $\Psi$, i.e. between branes and antibranes.

We can also give a physical interpretation to the canonical transformation $T^n$, which adds multiples of $p$ to $x$, $x \mapsto x-np$. Under this transformation, the constraint operator becomes
\ba
1 - e^{-p} - P(q) e^{np} e^{-x}   \nn
\ea
with the normal ordering ambiguity $P(q) = (-1)^n q^{\frac{1}{2}(1+2n)}$. It annihilates the partition function in framing $n$,
\ba
\Psi^{(n)}(x)&=&  \sum_{r=0}^{\infty} (-1)^r \frac{1}{[r]!} q^{-\frac{1}{4} (r-1)r(1+2n)} (-1)^{nr} e^{-rx} \,. \nn
\ea
The correspondence between the integer ambiguity arising by adding multiples of $p$ to $x$ and the framing ambiguity was already developed in \cite{AV}. It is gratifying to see this relation reproduced from the point of view pursued here.

\paragraph{The conifold:} Our starting point is $\IC \IP^3$ coordinatized by $(e^{-Y_1}: \cdots: e^{-Y_4})$, with the coordinates satisfying the constraint
\ba
Y_1 + Y_2 - Y_3 - Y_4 = t \nn  \,.
\ea
The equation for the Riemann surface characterizing the B-brane moduli space on the mirror of the conifold is given by
\ba
e^{-Y_1} + e^{-Y_2} + e^{-Y_3} + e^{-Y_4} = 0  \,.  \label{constraint conifold}
\ea
At each puncture on this surface, $\re Y_i \rightarrow \infty$ for two choices of $i$ (in terms of the A-model coordinates, these correspond to one base and one fiber coordinate of the resolved conifold). By our criteria, either choice qualifies as a canonical coordinate at the given puncture. Below, we will determine what this choice corresponds to by invoking the constraint equation. Upon making a choice, requiring the equation for the leaves of the foliation to be linear then fixes the patch from which our polarization must descend.

At $P_1$, e.g., $Y_2, Y_4 \rightarrow \infty$. The polarization for which $Y_2$ yields the canonical coordinate descends from the patch $y_3 \neq 0$, as the equation for the leaves of the foliation are then
\ba
\frac{y_1}{y_3} e^{-x} &=& \frac{y_4}{y_3} e^{-t} \,. \nn
\ea
The choice of sign of the symplectic form (\ref{symplectic form conifold}) then determines the appropriate sign of the canonical momentum, $e^{\pm p} = \frac{y_3}{y_1}$. We compile table \ref{concor} by choosing the canonical coordinate among the fiber coordinates for row I and the base coordinates for row II.

\begin{table}
\begin{scriptsize}
\begin{tabular}{c||c|c||c|c||c|c||c|c}
& \multicolumn{2}{c||}{$P_1:(1:0:-1:0)$} & \multicolumn{2}{c||}{$P_2:(1:0:0:-1)$} & \multicolumn{2}{c||}{$P_3:(0:1:0:-1)$}&\multicolumn{2}{c}{$P_4:(0:1:-1:0)$} \\ \cline{2-9}
 &$p_1$  &$x_1$ &$p_2$ & $x_2$ & $p_3$ & $x_3$ & $p_4$ & $x_4$ \\ \hline
 I &$ \log y_3/y_1$ & $-\log y_4/y_1$ & $-\log y_4/y_1$ & $-\log y_3/y_1$ & $\log y_4/y_2$& $-\log y_3/y_2$ & $ -\log y_3/y_2$ & $-\log y_4/y_2$ \\ \cline{2-9}
 & $p$ & $x$ & $x$ & $-p$ & $t-p$ & $-t-x$ & $-t-x$ & $-t+p$ \\ \hline
 II & $\log y_1/y_3$ & $-\log y_2/y_3$ & $-\log y_1/y_4$ & $-\log y_2/y_4$ & $ \log y_2/y_4$ & $-\log y_1/y_4$ & $ -\log y_2/y_3$ & $-\log y_1/y_3$ \\ \cline{2-9} 
 & $p$ & $x$ & $t-x$ & $t+p$ & $-t-p $ & $t-x $ & $x$ & $-p$ 
 \end{tabular} 
\end{scriptsize}
\caption{\small{Canonical variables for the conifold, with the canonical coordinate chosen along the fiber (I) or the base (II) \label{concor}.}}
\end{table}

Equation (\ref{constraint conifold}) expressed in terms of the canonical coordinates in the top row of table \ref{concor} takes the following forms,
\ba 
P_{1,3} &:& 1+ e^{-x}+e^{p}+e^{-x+p-t} = 0 \nn\\ 
P_{2,4} &:& 1+ e^{-x}+e^{-p}+e^{-x-p-t} = 0  \,.\label{rscon}
\ea
As the partition functions at $P_1, P_3$ and $P_2, P_4$ respectively that we determined in the previous section coincide, it is a first test on the consistency of our choice of canonical coordinates that we obtain the same equations at the respective pairs. And indeed, upon including the appropriate normal ordering constants, we find that the partition functions $\Psi^{1,2,3,4}$ are annihilated by the appropriate constraint operator,
\ba
(1- e^{p} - q^{-1/2} e^{-x} + q^{-1/2} e^{-x}Qe^{p}) \Psi^{1,3} &=& 0 \nn\\
(1- e^{-p} - q^{1/2} e^{-x} + q^{1/2} e^{-x}Qe^{-p}) \Psi^{2,4} &=& 0 \,. \nn
\ea
With the help of the constraint equation, we can determine the significance of the second choice of canonical coordinates. At a puncture, the base and fiber coordinate going to infinity satisfy, by the constraint equation, $Y_f^\infty - Y_b^0 = -t + Y_b^\infty - Y_f^0$, i.e. the canonical variable in the top row of table (\ref{concor}) is related to the variable in the bottom row via $x_{top}+t  =  x_{bottom}$. Together with $p_{top} = - p_{bottom}$, the equations (\ref{rscon}) are identical when expressed in terms of the top or bottom variables, up to the substitution $t \rightarrow -t$. The two choices of canonical variables hence correspond to the two choices of K\"ahler cone for the conifold.
In the sense that the relation $p_{top} = - p_{bottom}$ effects a change of phase space orientation, branes and antibranes are swapped under a flop transition.

The two operations $p \mapsto -p$ and $ (p,x)^T \mapsto T^n (p,x)^T$ have the same interpretation as in the case of $\IC^3$. That $p \mapsto -p$ distinguishes between branes and antibranes is immediate upon inspection of the two equations (\ref{rscon}): they are exchanged by $p \mapsto -p$, consistent with $\Psi^{1,3} = \Psi^{2,4*}$. The calculation establishing that acting with $T^n$ corresponds to a change of framing generalizes immediately from the $\IC^3$ case.

To summarize, upon choosing a sign for the symplectic form, the prescription of choosing the canonical variable to go to infinity at the puncture and the momentum to 0, supplemented for the conifold with the requirement that the leaves of the foliation be determined by a linear equation, fix an assignment of canonical coordinates to each puncture uniquely in the case of $\IC^3$, and up to a binary choice in the case of the conifold. This latter choice reflects the choice of K\"ahler cone.

In addition, we identify the physical significance of two operations
\begin{itemize}
\item $p \mapsto -p$ maps canonical coordinates suitable for describing branes to those suitable for describing antibranes.
\item $x \mapsto x + np$ maps to coordinates suitable for describing the open topological string at framing shifted by $n$ units.
\end{itemize}

In this subsection, evidence for the correct choice of canonical coordinates stemmed from application of the constraint equation. In the next subsection, we turn to the canonical transformations relating these coordinates, and their representations on the space of partition functions.

\subsection{Transformation of the partition function} \label{s:transform}
The $SL(2,\IZ)$ matrices $A_{ij}$ mapping $(p_j,x_j)$ into $(p_i,x_i)$ for $\IC^3$ and for the conifold can be read off from the tables \ref{preconcor} and \ref{concor} respectively. For $\IC^3$, they are  
\ba \nn
A_{12} = A_{23}= A_{31} &=&\left(
 \begin{matrix} 

      0 & -1 \\
      1 & -1 \\
   \end{matrix}
\right) = -TS \,,
\ea
where
\ba \nn
S=
\left(
 \begin{matrix} 
      0 & 1 \\
      -1 & 0 \\
   \end{matrix}
\right) \,, \hspace{1cm}
T=
\left(
 \begin{matrix} 
      1 & 0 \\
      1 & 1 \\
   \end{matrix}
\right) \,.
\ea
For the conifold, the transformation matrix from $P_2$ to $P_1$ is given by
\ba
A_{12} &=&\left(
 \begin{matrix} 
      0 & -1 \\
      1 & 0 \\
   \end{matrix}
\right) = -S\,, \nn
\ea
and the transformation between $P_1$ and $P_3$, $P_2$ and $P_4$ respectively is affine linear, i.e. $A_{ij} (p_j, x_j)^T + a_{ij} = (p_i, x_i)^T$, with
\ba \nn
A_{13} = A_{24} =\left(
 \begin{matrix} 
      -1 & 0 \\
      0 & -1 \\
   \end{matrix}
\right) =  S^2  \,,&& 
a_{13} = \left(
\begin{matrix}
t \\
-t \\
\end{matrix} \right) \,, \hspace{0.3cm}
a_{24} = \left(
\begin{matrix}
-t \\
-t \\
\end{matrix} \right)  \,.
\ea
\begin{figure}[h]
\psfrag{A12}{${\scriptstyle{-S}}$}
\psfrag{A13}{${\scriptstyle{S^2 \circ \,+ \,(t,-t)^{T}}}$}
\begin{center}
\epsfig{file=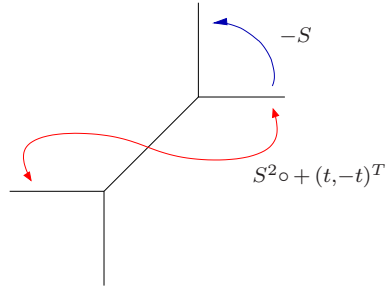, height=1.5in}
\end{center}
\caption{\small Canonical transformations relating different brane placements on the conifold.
\label{fig:cantransf}}
\end{figure}

We now want to study how the operators $T$ and $S$ are represented on the Hilbert space of which the open string partition function is to be a wave function. Having determined the canonical coordinates at the various punctures in the previous section, working out the transformation of a wave function under the respective change of polarization is a canonical exercise (see appendix \ref{QM}). The subtlety that arises for us however is that phase space is $M = \IC^* \times \IC^*$, i.e. the canonical variables are to be cylinder valued, while the symplectic form is of type (2,0). This invalidates the canonical treatment in the appendix (for which $X$ and $P$ are assumed to be self-adjoint). Pursuing the hybrid strategy suggested in the previous subsection yields suggestive results: following \cite{ADKMV}, we use the integration kernels obtained from the canonical approach,
\ba \label{intkernel}
K_A(x',x) &=& e^{\frac{1}{2g_s c} ( d x^2 - 2 xx' + a x'^2)}\,,
\ea
but give ourselves leeway in the choice of integration contour. Whenever $d\neq0$, we interpret the integral as a Gaussian with the contour chosen to ensure convergence, and evaluate
\ba \label{mastergauss}
\int K_A(x',x) e^{-nx} &\rightarrow& e^{(a-\frac{1}{d})\frac{x'^2}{2g_s c} - \frac{nx'}{d} - \frac{g_s c n^2}{2d}} \,.
\ea
With this understanding, we find that the action of the $T$ operator, which corresponds to a change of framing, can already be implemented at the level of the fermion operator, whereas our evaluation of the $S$ operator requires the specific choice of coefficients of the mode expansion which arise upon evaluating the VEV of the fermion operator.

The kernel that follows for the transformation $T^m$ is
\ba
K_{T^m}(x',x) &=& e^{\frac{1}{2g_s m}  (x^2 - 2 xx' +  x'^2)}\,. \nn
\ea 
The result of the integration term by term is
\ba \label{TK}
\int dx \, K_{T^m}(x',x) e^{-nx} &\sim& q^{-m \frac{n^2}{2}} e^{-nx'} \,. 
\ea
The coefficients of $e^{-nx}$ in the series expansion of $\Psi$ and $\Psi^*$ (for both the vertex and the conifold, with branes inserted on either leg) are shifted under a change of framing by $k$ units as
\ba
\Psi: a_{0,n-1} \mapsto (-1)^{kn} q^{-m \frac{n(n+1)}{2}} a_{0,n-1} \nn \\
\Psi^*: a_{n-1,0} \mapsto (-1)^{kn} q^{m \frac{n(n+1)}{2}} a_{n-1,0}\,. \nn
\ea
We can hence implement a framing shift by combining the integral transform (\ref{TK}) with a shift of the argument of the partition function,
\ba
\Psi^{(m)}(x') &=& \int dx \, K_{T^m}(x',x) \Psi(x + \frac{m}{2}g_s +m \pi i) \,,\nn\\
\Psi^{(-m)*}(x') &=& \int dx \, K_{T^m}(x',x) \Psi^*(x - \frac{m}{2}g_s +m \pi i) \,.\nn
\ea
Note that the integral kernel that induces a shift of framing by $m$ units for $\Psi$ induces a shift of $-m$ units for $\Psi^*$. This is consistent with our findings that the two partition functions are related by the opposite choice of sign of the symplectic form. This choice of sign manifests itself in the choice of sign of $\hbar$, which enters in the integral kernel multiplying $m$ (see appendix \ref{QM}).

Now let's turn to the transformation given by $S$. We read off the corresponding integration kernel from (\ref{intkernel}) to be
\ba
K_S(x', x) &=& e^{\frac{1}{g_s} x x'} \,.\nn
\ea
This results in a Fourier transform. We currently only know how to evaluate the transformation directly when the partition function is presented as an infinite product, i.e. for the particular choice of framing that yields $\Psi^{1,2,3,4}$ in the form given in (\ref{productrep24}) and (\ref{productrep13}). Here, we will choose a contour in the complex $x$ plane so that we pick up all residues at $x=g_s (i+\frac{1}{2})$, $i\ge0$. We then obtain for $\IC^3$
\ba
\oint \Psi(x) e^{\frac{1}{g_s}x x'} &=& 
\oint \prod_{i=0}^\infty \frac{1}{1-q^{i+\frac{1}{2}} e^{-x}} \,e^{\frac{1}{g_s}x x'} \nn\\
&=& \prod_{i=1}^{\infty} \frac{1}{1-q^i} \sum_{n=0}^{\infty} \prod_{i=1}^{n} \frac{1}{1-q^{-i}} e^{(n+\frac{1}{2})x'} \nn\\
&=& \frac{q^{-\frac{23}{24}}e^{\frac{1}{2}x'}}{ \eta(q)} \Psi^{*}(-x'-\frac{1}{2}g_s) \,,\nn
\ea
and likewise for the conifold,
\ba
\oint \Psi^{1,3}_t(x) e^{ x' x/g_s} &=& 
\oint \prod_{i=0}^\infty \frac{1- q^{i+ \frac{1}{2}} e^{-x}}{1-Q q^{i+\frac{1}{2}} e^{-x}} e^{x'x/g_s} \nn\\
&\rightarrow& (1-Q^{-1}) \prod_{i=1}^\infty \frac{1-Q^{-1} q^i}{1-q^{i}} \sum_{n=0}^{\infty} \prod_{i=1}^n \frac{1-Qq^i}{1-q^i} e^{(n+\frac{1}{2})x'}Q^{x'/g_s-n} \nn \\
&=&  \prod_{i=0}^\infty \frac{1-Q^{-1} q^i}{1-q^{i+1}} e^{\frac{g_s-2t}{2g_s} x'}  \Psi^{1,3*}_{t-g_s} (-x' -t +\frac{1}{2}g_s)
\ea
and
\ba
\oint \Psi^{2,4}_t(x) e^{x' x/g_s} &=& 
\oint \prod_{i=0}^\infty \frac{1- Qq^{i+ \frac{1}{2}} e^{-x}}{1-q^{i+\frac{1}{2}} e^{-x}} e^{ x' x/g_s} \nn\\
&\rightarrow& (1-Q) \prod_{i=1}^{\infty} \frac{1-Qq^i}{1-q^i} \sum_{n=0}^{\infty} \prod_{i=1}^{n} \frac{1-Qq^{-i}}{1-q^{-i}} e^{(n+\frac{1}{2})x'} \nn\\
&=&  \prod_{i=0}^{\infty} \frac{1-Qq^i}{1-q^{i+1}}  e^{\frac{1}{2}x'} \Psi^{2,4*}_{t+g_s}(-x'-\frac{1}{2}g_s) \,.\nn
\ea
Note that $g_s$ shifts of closed string K\"ahler parameters have already made several appearances in studies of the open topological A-model, starting with \cite{DF}.

The integral kernel for the transformation $S^2=-1$ cannot be read off of (\ref{intkernel}), as $c=0$. It is not hard however to convince oneself that the correct kernel here is simply
\ba
K_{S^2}(x', x) &=& \delta(x + x') \,.\nn
\ea
Let's now check whether these results are consistent with the canonical coordinates determined at each puncture.

\paragraph{\mbox{\boldmath$\IC^3$}:} The transformation from $P_2$ to $P_1$ for $\IC^3$ is given by $S^2 TS$. Ignoring the coefficients and the shifts of the argument by $g_s$, the corresponding integral transformations act as 
\ba
\Psi(x) \xrightarrow{S} \Psi^*(-x) \xrightarrow{T} \Psi(-x) \xrightarrow{S^2} \Psi(x) \,. \nn
\ea 
Aside from the various shifts and overall coefficients, the partition functions at the punctures $j$ are hence indeed related to those at $i$ by the $S^2 T S$ integral transform (with $(i,j)\in \{(1,2),(2,3),(3,1)\}$, as throughout). Instead of performing the transformations one by one, we can simply invoke (\ref{mastergauss}), 
as unlike $S$, $S^2 TS$ has $d \neq 0$. The result is
\ba \label{vtran}
\Psi(x + \frac{1}{2}g_s + \pi i) \xrightarrow{S^2 T S} e^{\frac{x^2}{2 g_s}} \Psi^*(-x)  \,. 
\ea
To rewrite the transform in a power series that converges for large $x$, note first that the partition function on $\IC^3$ satisfies the following identity (which is essentially the Jacobi triple product identity, see appendix \ref{hypermod}),
\ba
\Psi^*(-x) &=& \Psi(x) \frac{1}{(q;q)_\infty} \sum_{n=-\infty}^\infty (-1)^n q^{n^2/2} e^{-n x}  \nn \\
&=& \Psi(x) \Phi(\tau, \zeta)  \,,  \label{modular}
\ea
with $q=e^{2\pi i \tau}$ and $x=-2\pi i\zeta$. $\Phi(\tau,\zeta)$ behaves well, upon continuation to complex $\tau$, under modular transformations. After the transformation, we can take the $g_s \rightarrow 0$ limit, obtaining $\Phi(\tau,\zeta) \rightarrow e^{-\frac{(x-i\pi)^2}{2 g_s}} $.  We are dropping an infinite sum in $e^{-1/g_s}$ corrections. We would like to interpret these terms as a sign for non-perturbative terms we are missing in the partition function.\footnote{We thank R. Dijkgraaf for first suggesting this interpretation.} We will comment on this further in the next subsection. With this interpretation, we can now complete (\ref{vtran}) to the following
\ba
\Psi(x + \frac{1}{2}g_s + \pi i) \xrightarrow{S^2 T S} e^{\frac{x^2}{2 g_s}} \Psi^*(-x)  \rightarrow  e^{\pi i \frac{x}{g_s} + \frac{\pi^2}{2 g_s}} \Psi(x)\,, \nn
\ea
which coincides with the result of applying the transformations $S$, $T$, $S^2$ consecutively. 

\paragraph{The conifold:} The transformation from $P_1$ to $P_3$ and from $P_2$ to $P_4$ is given by $S^2=-1$ and shifts by $t$. Our transformation rules hence imply that the partition functions at the points of the two pairs should be related, respectively, by replacing the argument $x \mapsto -x -t$.
From our calculations above, the partition functions within each pair are equal. To check whether these two results are compatible, we must hence relate the one point function at $x$ to the one point function at $-x$. To this end, note that at the points $P_2$ and $P_4$, the partition functions for the conifold can be expressed in terms of the partition function of $\IC^3$ as follows,
\ba
\Psi^{2,4} (x) &=& \frac{\Psi(x)}{\Psi(x+t)} \,. \nn
\ea
Using (\ref{modular}), we thus obtain
\ba
\Psi^{2,4} (-x-t) &=& \frac{\Psi(-x-t)}{\Psi(-x)} \nn \\
&=& \frac{\Psi(x)}{\Psi(x+t)} \frac{\Phi(\tau, \zeta)}{\Phi(\tau, \zeta -\frac{t}{2\pi i})} \nn\\ 
&\rightarrow&  e^{-\frac{1}{2g_s}[t^2 + 2t (x-i \pi)]} \, \Psi^{2,4}(x)  \label{ancontp2} \,.
\ea
Likewise for $\Psi^{1,3}$, 
\ba
\Psi^{1,3} (-x-t)  \rightarrow e^{\frac{1}{2g_s}[t^2 + 2t (x-i \pi)]} \,  \Psi^{1,3} (x)  \,.\nn
\ea
Up to the coefficient and, again, $\cO(e^{-1/g_s})$, we conclude that the transformations are consistent with our choice of canonical coordinates. 

The transformation for the conifold from $P_2$ to $P_1$ and $P_4$ to $P_3$ is given by $- S$. It acts as
\ba
\Psi^{2,4}(x) \xrightarrow{S} \Psi^{2*,4*}(-x) = \Psi^{1,3}(-x) \xrightarrow{S^2} \Psi^{1,3}(x) \,. \nn
\ea
With the same caveats as above, the transformation kernel relating $\Psi^{2,4}$ to $\Psi^{1,3}$ hence follows correctly from the canonical coordinates assigned to $P_{1,3}$ and $P_{2,4}$. The transformation from $P_{1,3}$ to $P_{2,4}$ is given by $(-S)^{-1} = S$. Here,
\ba
\Psi^{1,3}(x) \xrightarrow{S} \Psi^{1*,3*}(-x-t) =  \Psi^{2,4}(-x-t) \sim \Psi^{2,4}(x) + \cO(e^{-1/g_s}) \,, \nn
\ea
where $\sim$ indicates that we have dropped the coefficient in (\ref{ancontp2}).

\subsection{Non-perturbative terms: a preliminary study} \label{s:non-per prelim}
What is the nature of the missing $\cO(e^{-1/g_s})$ contributions? If the Chern-Simons proposal of \cite{cswitten} were the last word on the target space description of the open topological A-model, we would expect this description (at least in cases where the worldsheet instanton contribution to the Chern-Simons action is convergent) to provide a non-perturbatively complete description of the theory. We have witnessed throughout our analysis however that this description must be modified: a complexified version of the theory seems to be required. We emphasize again the distinction to complex Chern-Simons theory; the required modification has the same number of fields as real Chern-Simons theory, and coincides with it perturbatively. In this subsection, we find a further indication pointing towards the necessity for such a modification.

In our study so far, we have followed the literature in assuming\footnote{In this subsection, we use conventions such that $x \rightarrow x + 2\pi$ is the periodicity of the exponential; the substitution $x \mapsto ix$ reinstates the conventions used in the rest of the paper. We will also express our results in terms of the Chern-Simons coupling $k$, rather than $g_s$. The two are related via $g_s = -\frac{2 \pi i}{k}$.}
\ba
\frac{\langle e^{ix} | e^{\oint_r A} | \Psi \rangle}{\langle e^{ix}  | \Psi \rangle  } &=& e^{irx} \,.  \label{lit}
\ea
On general grounds, normalized expectation values of Wilson loops in Chern-Simons theory on solid tori are known to yield characters of the relevant affine Lie algebra, in our case $\mathfrak{u}_k$. What has been neglected in (\ref{lit}) is to mod out by large gauge transformations (see \cite{marino}, which argues for this approach). If we instead take these into account, (\ref{lit}) is modified to \cite{Elitzur}
\ba
\Psi_r(x) &=&  \exp[i r x  ]  \sum_{m \in \IZ}  \exp[ik m x] \,, \label{modified}
\ea
which is indeed a $\mathfrak{u}_k$ character.

How does this modification effect our considerations? Let us revisit the transformation of $\Psi^*(x)$ in the case of $\IC^3$. Recall that the canonical transformation $-TS$ previously yielded, schematically,
\ba
\Psi^*(x) \mapsto_{-TS} \Psi(-x)  \,.
\ea
By 
\ba
\Psi^*(x) \Psi^*(-x) \sim \frac{\vartheta(x,1/k)}{\eta(1/k)}  \,,
\ea
we argued for $\Psi(-x) \sim \Psi^*(x) + \cO(e^{-1/g_s})$.

The wave function for the antibrane with the modification (\ref{modified}), which we denote by $\tilde{\Psi}^*$, is given by
\ba
\tilde{\Psi}^*(x) &=& \sum_{r=0}^{\infty} \frac{1}{[r]!} q^{\frac{1}{4}(r-1)r} e^{irx} \sum_{m \in \IZ} e^{ikmx} \,.
\ea

Evaluating the Gaussian integral associated to the canonical transformation $-TS$ for the modified wavefunction term by term yields
\ba
\int dx \,K_{-TS} (x',x) e^{ix (r + km)} &=& e^{\frac{x'^2}{2 g_s} - i(r+km)x' + \frac{g_s (ir + ikm)^2}{2}} \\
&=& e^{-\frac{k x'^2}{4 \pi i}} e^{-i(r+km) x'} q^{-\frac{r^2}{2}} e^{\pi i  m^2 k} \,,
\ea
hence formally,
\ba
\tilde{\Psi}^*(x) \mapsto_{-TS}& & \Psi(-x-\pi) \, e^{-\frac{kx^2}{4 \pi i}} \vartheta(-\frac{kx}{2\pi},k)  \nn \\
&=& \sqrt{i/k} \Psi(-x-\pi)   \vartheta(\frac{x}{2 \pi},  -1/k) \nn \\
&=& \sqrt{i/k} \;\eta(-1/k) q^{-1/24}  \; \Psi^*(x) \,. \label{modtrans}
\ea
We see that the modification (\ref{modified}) yields precisely the $\vartheta$ function needed to relate $\Psi(-x)$ to $\Psi^*(x)$. Two main points remain to be understood:
\begin{itemize}
\item{The coefficient in (\ref{modtrans}), as well as the fact that we obtain $\Psi^*(x)$, and not $\tilde{\Psi}^*(x)$, are possibly due to our considering a too na\"ive measure in performing canonical transformations (see \cite{Elitzur} for a discussion of the measure factor in the non-abelian context).}
\item{Our considerations regarding theta functions are formal as real $k$ corresponds to real modular parameter $\tau$, and this a degenerate limit in which theta functions are ill-defined. We take this as yet another indication that a complex variant of the conventional Chern-Simons theory is needed to capture the open topological string.}
\end{itemize}
We are currently investigating these matters, and hope to report on our results elsewhere.

\subsection{Transformation of the two point function} \label{s:transop}
As we saw above, the open string partition function is given by the one point function of the fermion operator. One can ask \cite{ADKMV} whether the transformation properties of the partition function studied in the previous subsection can be attributed to the fermion operator. If so, one would expect to be able to deduce the relation between the higher point operators as well. 

Following \cite{ADKMV}, we look to the tranformation properties of the brane/antibrane correlator to study this question. The equality under examination is
\ba \label{2pttrans}
\langle 0 | \psi^i(y) \psi^{j*}(x) | Z \rangle &\overset{?}{=}& \int d\tilde{y} \,K_{A_{ij}}(y,\tilde{y})\langle 0 | \psi^j(\tilde{y}) \psi^{j*}(x) | Z \rangle  \,.
\ea
By the identity (\ref{idbraneantibrane}), which hold for $\IC^3$ \cite{ADKMV} as well as for the conifold, the RHS of (\ref{2pttrans}) is given by
\ba
 \int d\tilde{y} \,K_{A_{ij}}(y,\tilde{y})\langle 0 | \psi^j(\tilde{y}) \psi^{j*}(x) | Z \rangle  &=& \int d\tilde{y} \,K_{A_{ij}}(y,\tilde{y}) \frac{\Psi^j(\tilde{y})}{e^{\tilde{y}} - e^{x}}  \Psi^{j*}(x) \,, \nn
\ea
while the LHS is 
\ba
\langle 0 | \psi^i(y) \psi^{j*}(x) | Z \rangle &=& \sum_{m,n=0}^{\infty} a_{mn}^{ji} e^{-(m+1)x} e^{-(n+1)y} \,. \nn
\ea
The identity (\ref{2pttrans}) hence boils down to a relation between the coefficients $a_{mn}^{ji}$ and the products $a_{r0}^i a_{0s}^j$.
We will consider the calculation for the vertex and for the conifold in turn.

\paragraph{$\IC^3$:} From (\ref{vertex12}), we read off
\ba
(-1)^{n} a^{ji}_{mn} &=& C^{ji} \sum_{k=0}^{\min(m,n)} (0| m-k-1)(0|n-k-1) q^{\frac{1}{2} \hkappa(m)}  \,, \nn\\
&=& C^{ji} (-1)^{m+n} \sum_{k=0}^{\min(m,n)} a_{0,m-k-1}^{jj} a_{0,n-k-1}^{ii} q^{\frac{1}{2} \hkappa(m)}  \,, \label{ijtoii}
\ea
where we have set $-a^{ii}_{0,-1} = a^{jj}_{-1,0} =1$.
Hence,
\ba  \label{2ptstraight}
\langle 0 | \psi^i(y) \psi^{j*}(x) | Z \rangle &=& C^{ji} \Psi^i(y) \sum_{k=0}^\infty (-1)^k \Psi^{j*}(x- g_s(k+1)) q^{\frac{1}{2}k(k+1)} e^{-(k+1)(x+y)}  \,.\nn \\
\ea
Now consider
\ba
\int d\tilde{x}\, K_{A^*_{ji}}(x,\tilde{x}) \Psi^i(y) \frac{\Psi^{i*}(\tilde{x})}{e^y - e^{\tilde{x}}} \nn \,,
\ea
with
\ba
A^*_{ji}  &=&\left(
 \begin{matrix} 
      -1 & -1 \\
      1 & 0 \\
   \end{matrix}
\right)  \,. \nn 
\ea
Since this transformation matrix has $d=0$, we cannot perform a Gaussian integration along the lines of (\ref{mastergauss}). We can try to bypass this difficulty, following the authors of \cite{ADKMV}, by decomposing the transformation as $A^*_{ji} = A^*_{jk} A^*_{ki}$, with 
\ba
A^*_{jk} = A^*_{ki} &=&\left(
 \begin{matrix}
      0 & 1 \\
      -1 & -1 \\
   \end{matrix}
\right) \,. \nn
\ea
Then
\ba
\int d\tilde{x}\, K_{A^*_{ki}}(x,\tilde{x}) \frac{\Psi^{i*}(\tilde{x})}{e^y - e^{\tilde{x}}} &=& -e^{-\frac{1}{2}x} q^{\frac{1}{8}} \sum_{k=0}^{\infty} e^{-(k+1)y} \Psi^{k*}(x+i\pi +g_s (k-\frac{1}{2})) q^{-\frac{1}{2}k}   \nn \,,
\ea
and
\ba
\lefteqn{q^{\frac{1}{8}} \sum_{k=0}^{\infty} e^{-(k+1)y} q^{-\frac{1}{2}k} \int d\tilde{x}\, K_{A^*_{jk}}(x,\tilde{x}) \,e^{-\frac{1}{2}\tilde{x}}\, \Psi^{k*}(\tilde{x}+i\pi+g_s (k-\frac{1}{2}))=} \nn \\
 &=& - \sum_{k=0}^{\infty} e^{-(k+1)(x+y)} \Psi^{j*}(x-g_s (k+1)) q^{\frac{1}{2}k(k+1)} e^{\frac{1}{2}(x+g_s)} \nn \,,
\ea
where we have used (\ref{modular}) repeatedly. Up to coefficients, this calculation thus reproduces (\ref{2ptstraight}). We now turn to the analogous computation for the conifold.

\paragraph{The conifold} $a^{13}$ and $a^{24}$ prove to satisfy a relation similar to (\ref{ijtoii}). We derive the following relation in the appendix,
\ba
a_{m n}^{42} &=& -C^{42} \sum_{k=0}^{\min(m,n)}  a^{44}_{m-k-1,0} \,a^{22}_{0,n-k-1} Q^{k}  \,.\nn
\ea
Hence,
\ba
\langle 0 | \psi^2 (y) \psi^{4*}(x) | Z \rangle &=&
\sum_{m,n=0}^{\infty} a_{mn}^{42} e^{-(m+1)x} e^{-(n+1)y} \nn \\
&=& -C^{42} \,\Psi^2(y) \Psi^{4*}(x) \sum_{k=0}^\infty e^{-(k+1)(x+y)} Q^{k} \nn \\
&=& - C^{42}\, e^{-x} \, \frac{\Psi^2(y) \Psi^{4*}(x)}{e^{y}-e^{-x-t}} \,. \label{2star4}
\ea
On the other hand,
\ba
\int d\tilde{y} \,K_{A^*_{42}}(x,\tilde{x}) \langle 0 | \psi^2(y) \psi^{2*}(\tilde{x}) \rangle &=& \int d\tilde{y} \,\delta(x+ \tilde{x} +t) \frac{\Psi^2(y) \Psi^{2*}(\tilde{x})}{e^{y} - e^{\tilde{x}}} \nn \\
&\sim&   \frac{\Psi^2(y) \Psi^{4*}(x)}{e^{y} - e^{-x-t}} \,.
\ea
We have here used $(\ref{ancontp2})$ and omitted the $t$ dependent prefactor arising upon the analytic continuation. Up to prefactors, this result agrees with (\ref{2star4}).

The two examples considered support the proposal of \cite{ADKMV} that the fermion operators themselves transform as wave functions under canonical transformations. We should point out however that we have made choices along the way (in particular the choice of examples). We interpret our results for the fermion operator as indicative of interesting structure to be explored. 

\section{Conclusion}
We have found strong indications that the open topological string partition function on the conifold behaves as a wave function. This is consistent with the existence of a target space description of the topological string, given the B-model result that the moduli space of branes on the conifold is connected. At various points in our analysis, we have encountered indications that the correct non-perturbative physics of the topological string should be captured by a complexified version of Chern-Simons theory. These include the need for a non-standard symplectic form of degree (2,0), canonical transformations being implemented by contour integrals, and the incorporation of non-perturbative terms seeming to favor complex Chern-Simons coupling $k$. Finding an explicit formulation of this complexified theory is an intriguing open problem. A better understanding of this theory should also allow a {\it derivation} of the prescription we put forth, extending the proposal of \cite{ADKMV}, for assigning canonical coordinates to punctures of the moduli space of branes.

We have demonstrated that the simple structure of the vacuum in terms of fermion bilinears verified in \cite{ADKMV} for $\IC^3$ persists for the conifold. Regarding wave function behavior, our results are suggestive of a further layer of structure associated with the fermion operators on the brane moduli space which promises to reward further study.

On a less technical note, it is intriguing, as already pointed out by Witten \cite{witten} long ago, that the worldsheet approach to the topological string makes some of the choices required for defining the path integral from the point of view of the target space theory. With an eye towards extracting lessons for the full physical string, this is an observation worth keeping in the back of ones mind.

\section*{Acknowledgements}
We would like to thank Jan de Boer, Robbert Dijkgraaf, Sheer El-Showk, Bartomeu Fiol, Lotte Hollands, Marcos Mari\~no, Martin Ro\v{c}ek, and Eric Verlinde for very helpful discussions and useful suggestions, and Robbert Dijkgraaf for reading the manuscript.

This work was supported by Stichting FOM.

\appendix
\section{Transformations of wave functions in quantum mechanics} \label{QM}
Given a state $|\psi \rangle$ and two bases $|x\rangle$ and $|x' \rangle$, the wave function representations of $| \psi \rangle$ are related by
\ba
\langle x' | \psi \rangle &=& \int \langle x' | x \rangle \langle x | \psi \rangle \,. \nn
\ea
Let us determine the matrix element $\langle x' | x \rangle$ between two bases related by a linear canonical transformation
\ba
P' &=& a P + b X \nn\\
X' &=& c P + d X  \,. \nn
\ea
We are considering the conventional case of real (self adjoint) $X$ and $P$, i.e.  $[X,P]=i\hbar$, realized by $X$ a multiplication operator, $P = -i\hbar \frac{\partial}{\partial x}$ and $\langle x | p \rangle = e^{\frac{i}{\hbar}px}$. It is easy to derive the following differential equation for the matrix element,
\ba
(x'-dx) \la x | x' \ra &=& -i\hbar c \frac{\partial}{\partial x} \la x | x' \ra \,,\nn
\ea
with the solution
\ba
\la x | x' \ra &=& e^{\frac{i}{\hbar c} (x x' - \frac{d x^2}{2}) + \gamma} \,,\nn
\ea
and $\gamma$ an integration constant.
Inverting the symplectic transformation, we obtain
\ba
\la x' | x \ra &=& e^{-\frac{i}{\hbar c} (x x' - \frac{a x'^2}{2}) + \gamma'} \,.\nn
\ea
Imposing hermiticity of the matrix element then yields
\ba 
\la x | x' \ra &=& e^{\frac{i}{2\hbar c} (- d x^2 + 2 xx' - a x'^2)}\,.\nn
\ea
We will set $\hbar = ig_s$, such that $[P,X]=g_s$. For $g_s$ real, this is of course not in accord with self-adjointness of the operator $Y$. Replacing the expansion variable $q^{g_s}$ with $q^{ig_s}$ on the other hand corresponds to the choice $\hbar = -g_s$.
  
\section{The mirror geometries} \label{s:tmg}
In the Hori-Vafa \cite{HV} prescription, the superpotential of the mirror Landau-Ginzburg theory to the A-model on a non-compact toric target space, presented as in (\ref{setup}) above, is given by 
\ba
W &=& \sum_{i=1}^n e^{-Y_i} \,, \nn
\ea
with $\re Y_i = |X_i|^2$, and $n-3$ linear constraints among the $Y_i$ reflecting the toric data. In \cite{HIV}, it is shown that as far as holomorphic data is concerned, we can instead consider the B-model on the geometry
\ba
\sum_{i=1}^n e^{-Y_i} &=&uv \,,\nn
\ea
with $u$ and $v$ sections of the appropriate line bundle. Non-compact  holomorphic cycles in this geometry that can be wrapped by B-branes are given by 
\ba
uv &=& c \nn
\ea 
for some $c \in \IC$. A distinguished class amongst these is given at $c=0$, as these do not exhibit a deformation corresponding to changing $c$. \cite{AV} identify these as the mirrors to the Lagrangian branes ending on the toric skeleton, discussed in section (\ref{setup}).
The moduli space of these branes is given by the Riemann surface
\ba
\sum_{i=1}^n e^{-Y_i} &=& 0 \,, \nn
\ea
which we can think of as cut out of $\IC \IP^n$ by this and the $n-3$ constraint equations, these being homogeneous in the coordinates $y_i = e^{-Y_i}$.

For $\IC^3$, the above prescription yields the 3-punctured sphere, given by the following linear equation in $\IP^2$,
\ba
e^{-Y_1} + e^{-Y_2}+e^{-Y_3} = 0  \,. \nn
\ea
The three punctures lie at $(0:1:-1)$, $(1:0:-1)$, and $(1:-1:0)$. They correspond to going off to infinity along the three edges of the toric diagram.

For the conifold, the following equation arises
\ba \label{eqrs}
e^{-Y_1}+e^{-Y_2}+e^{-Y_3}+e^{-Y_4}=0 \,,
\ea 
with the constraint
\ba  \label{constraint}
Y_1 + Y_2 - Y_3 -Y_4 = t \,.
\ea
In terms of the homogeneous coordinates $y_i = e^{-Y_i}$ on $\IP^3$, we obtain the following two equation of degree 2,
\ba
y_1 y_3 + y_2 y_3 + y_3^2 + e^t y_1 y_2 &=& 0 \,, \nn \\
y_1 y_2 - e^t y_3 y_4 &=& 0 \,. \nn
\ea
The punctures lie at $P_1 : (1:0:-1:0)$, $P_2:(1:0:0:-1)$, $P_3:(0:1:0:-1)$, $P_4:(0:1:-1:0)$. At each puncture, the real part of two of the coordinates $Y_i$ is going off to infinity.

In the $Y_i$ coordinates, the 4-punctured sphere naturally appears as 4 cylinders which are joined. As promised in the text, the moduli space of branes that arises in the B-model setup hence appears to connect the disconnected moduli space that arose in the A-model picture.
  
\section{Assorted comments on representations} \label{assorted}
The irreducible representations of the symmetric group $S_n$ are given by partitions of $n$, i.e. by Young tableaux with $n$ boxes. The trivial representation is the tableau $(n-1|0)$, the standard representation is $(n-2|1)$, and its $d^{th}$ exterior power is $(n-d-1|d)$ (see appendix \ref{apschur} for the notation regarding Young tableaux). Representations of $GL(V)$ are constructed from $V$ by application of the Weyl functor, which maps $V$ to the image of the Young symmetrizer $c_\lambda \in \IC S_d$ acting on $V^{\otimes d}$. This image is empty when the number of rows of the Young tableau exceeds the dimension of $V$. The fundamental representation is hence given by $(0|0)$, the $d^{th}$ symmetric product by $(d-1|0)$, etc.

Frobenius' formula (see e.g. chapter 4 of \cite{fultonharris} for an introduction) has as input the class ${\bf k}$ of a cycle of $S_d$, i.e. $\sum i k_i = d$, and yields a polynomial, the respective coefficients of which correspond to the character of elements in the class $C_{{\bf k}}$ in the various representation. From this starting point, one can derive a polynomial that depends on the representation, and characters are obtained by evaluating the polynomial at $x_i = k_i$. This formula can also be written as
\ba
\prod_{i=1}^d P_i(x)^{k_i} &=& \sum_{|\lambda|=d,\,\lambda_{n+1}=0} \chi_{\lambda}(C_{\bf k}) s_\lambda  \,,\nn
\ea
where $P_i(x) = x_1^i + \ldots + x_n^i$. Introducing the $n \times n$ matrix $V$ with eigenvalues $x_1, \ldots, x_n$, this is
\ba
\prod_{i=1}^d (\tr V^i)^{k_i} &=& \sum_{|\lambda|=d,\,\lambda_{n+1}=0} \chi_{\lambda}(C_{\bf k}) \Tr_{\!\! \lambda} V \,.\nn
\ea
We have here used the fact that the Schur polynomial $s_\lambda$ is the character of the $GL(n)$ representation given by $\lambda$.

\section{Basic hypergeometric series, the $\vartheta$-function, and the open topological string partition function} \label{hypermod}
The basic hypergeometric series ${}_r \phi_s$ is defined as \cite{gr}
\ba
{}_r \phi_s (a_1, \ldots, a_r; b_1, \ldots, b_s;q,z) &:=& \sum_{n=0}^{\infty} \frac{(a_1;q)_n \cdots (a_r;q)_n}{(q;q)_n (b_1;q)_n \cdots (b_s;q)_n} \left[(-1)^n q^{\frac{n(n-1)}{2}} \right]^{1+s-r} z^n \nn \,,
\ea
where
\ba
(a;q)_{n} = \prod_{i=0}^{n-1} (1-aq^i) \,.\nn
\ea
Note that in the notation used in the text,
\ba
(q;q)_n = (-1)^n [n]! q^{\frac{n(n+1)}{4}}  \,.\nn
\ea
${}_0 \phi_0$ has a representation as an infinite product,
\ba
{}_0 \phi_0 &=& \sum_{n=0}^\infty (-1)^n \frac{q^{n(n-1)/2}}{(q;q)_n} z^n  \nn \\
&=& \prod_{n=0}^{\infty} (1-zq^n) \nn \\
&=:& (z;q)_{\infty} \nn \,.
\ea
Using the ratio test, we can easily check that this series is absolutely convergent for $q < 1$ and $z \in \IC$. In this notation, the Jacobi triple product identity takes the form
\ba
(zq^{1/2};q)_\infty \,(\frac{q^{1/2}}{z};q)_\infty \,(q;q)_\infty = \sum_{n=-\infty}^\infty (-1)^n q^{n^2/2} z^n \,.\nn
\ea
In terms of basic hypergeometric series, the partition function $(\ref{oneptexpstar})$ at $Q=0$ can be expressed as
\ba
\Psi^*(z)&=& \sum_{n=0}^{\infty}  \frac{1}{[n]!} q^{\frac{1}{4} (n-1)n} z^n \nn \\
&=& \sum_{n=0}^{\infty} \frac{(-1)^n}{(q;q)_n} q^{\frac{1}{4} n(n+1)}q^{\frac{1}{4} (n-1)n} q^{-\frac{n}{2}} (q^{\frac{1}{2}} z)^n \nn\\
&=& (zq^{1/2};q)_\infty \,. \nn
\ea
The Jacobi triple product identity hence implies the following relation between the one point functions,
\ba \label{jacobi}
\Psi^*(z) \Psi^*(\frac{1}{z}) =  \frac{1}{(q;q)_\infty} \sum_{n \in \IZ} (-1)^n q^{n^2/2} z^n \,. 
\ea
The RHS of the above equation can be expressed in terms of two modular forms (see e.g. \cite{mumford}), the $\eta$ and $\vartheta$ function,
\ba
(q;q)_\infty &=& \prod_{n=0}^\infty ( 1 - q^{n+1}) \nn\\
&=& q^{-\frac{1}{24}} \eta(\tau)  \,,\nn
\ea
and 
\ba
\vartheta(\zeta,\tau) &=& \sum_{n \in \IZ} e^{\pi i n^2 \tau + 2 \pi i n \zeta} \nn\\
&=& \sum_{n \in \IZ} q^{n^2/2} z^n \,,\nn
\ea
with $q=e^{2 \pi i \tau}$ and $z=e^{2 \pi i \zeta}$, i.e. $g_s = 2 \pi i \tau$, $x= - 2\pi i \zeta$.
We want to interpret the RHS of (\ref{jacobi}) as instanton corrections that can be dropped in perturbation theory. To this end, we consider the modular transformations for which $g_s \rightarrow  \frac{1}{g_s}$, 
\ba
\eta(-1/\tau) = \sqrt{-i\tau} \eta(\tau) \nn
\ea
and
\ba
\vartheta(\zeta/\tau, -1/\tau) &=& \sqrt{-i \tau} e^{\pi i \zeta^2/\tau} \vartheta(\zeta,\tau)   \nn \,,
\ea
thus obtaining
\ba
 \frac{1}{(q;q)_\infty} \sum_{n=-\infty}^\infty (-1)^n q^{n^2/2} z^n &=& \frac{q^{\frac{1}{24}}}{\eta(\tau)} \vartheta(\zeta - \frac{1}{2},\tau)  \nn\\
 &=& \frac{q^{\frac{1}{24}}}{\eta(-1/\tau)} e^{-\pi i( \zeta + \frac{1}{2})^2/\tau} \vartheta( \frac{\zeta + \frac{1}{2}}{\tau}, -\frac{1}{\tau}) \,. \nn
\ea
The exponential coefficient expressed in terms of $g_s$ and $x$ is
\ba
e^{-\pi i( \zeta + \frac{1}{2})^2/\tau} &=&  e^{-\frac{(x-i\pi)^2}{2 g_s}}  \,.\nn
\ea

\section{Identities and computations involving Schur functions} \label{apschur}
For $x_i = q^{\rho_i}$, $\rho_i = i-1$, $i=1, \ldots$, \cite{macdonald} obtains the following expression for the Schur function,
\ba
s_\lambda &=& q^{n(\lambda)} \prod_{x\in \lambda} (1 - q^{h(x)})^{-1} \nn\,,
\ea
where $n(\lambda) = \sum_{i \ge 1} (i-1) \lambda_i$ and $h(x)$ is the hook length of $x\in \lambda$. With $\rho_i = -i+\frac{1}{2}$, $i=1, \ldots$, we obtain instead
\ba \label{schur}
s_\lambda &=& q^{-n(\lambda)} \prod_{x\in \lambda} (1 - q^{-h(x)})^{-1} q^{-|\lambda|/2} \,. 
\ea
From now on, and in the body of the paper, we will use the same symbol for a representation, the associated Young tableaux, and the associated Schur function with arguments $x_i = q^{-i + \frac{1}{2}}$. 
The topological vertex also involves Schur functions $s_\lambda$ with arguments $x_i = q^{\alpha_i - i +\frac{1}{2}}$ for a Young tableaux $\alpha$ with $\alpha_i$ boxes in the $i^{th}$ row. We denote these as $\lambda(\alpha)$. Skew Schur functions $s_{\lambda/\mu}$ with the respective arguments are denoted as $\frac{\lambda}{\mu}$ and $\frac{\lambda}{\mu}(\alpha)$.

We denote Young tableaux by indicating the number of boxes (to the right $|$ below) the diagnonal, as $(m_1, \ldots, m_r | n_1, \ldots, n_r)$, with $m_1 > \ldots > m_r$, $n_1 > \ldots > n_r$. For hook representations, this reduces to $(\alpha_1|\alpha_2)$, with $\alpha_1$ boxes to the right and $\alpha_2$ boxes below the diagonal (notice that the total number of boxes is hence $\alpha_1+\alpha_2+1$). Specializing (\ref{schur}) to hook representations yields
\ba
(\alpha_1|\alpha_2) &=&q^{-\frac{\alpha_1+\alpha_2+1}{2}}q^{-\sum_{i=1}^{\alpha_2} 1} (1-q^{-(\alpha_1+\alpha_2+1)})^{-1}\prod_{i=1}^{\alpha_1}(1-q^{-i})^{-1} \prod_{j=1}^{\alpha_2} (1- q^{-j})^{-1} \nonumber \\
&=& q^{-\frac{\alpha_2(\alpha_2+1)}{2}} [\alpha_1+\alpha_2+1]^{-1}\prod_{i=1}^{\alpha_1}q^{\frac{i}{2}} [i]^{-1} \prod_{j=1}^{\alpha_2} q^{\frac{j}{2}}[j]^{-1} \nn \\
&=&  \frac{q^{\frac{\kappa(\alpha_1|\alpha_2)}{4}}}{[\alpha_1]![\alpha_2]![\alpha_1+\alpha_2+1]} \,, \label{schurhook}
\ea
where we have introduced
\ba
[\alpha] = q^{\frac{\alpha}{2}} - q^{-\frac{\alpha}{2}}\,\, \,, \hspace{1cm}[\alpha]! = \prod_{\beta=1}^{\alpha} [\beta] \,, \nn 
\ea
and  
\ba
\kappa(\alpha_1|\alpha_2) &=& \alpha_1(\alpha_1+1) - \alpha_2 (\alpha_2+1) \,. \nn
\ea
We will also find the following notation useful,
\ba
\hkappa(\alpha_1) &=& \alpha_1 (\alpha_1+1) \nn \,.
\ea
To evaluate skew Schur functions for hook representations, we first apply the Littlewood-Richardson rules to obtain, for $\alpha_1 \neq \eta_1$ and $\alpha_2 \neq \eta_2$,
\ba
\frac{(\alpha_1|\alpha_2)}{(\eta_1|\eta_2)} &=& (\alpha_1-\eta_1|\alpha_2-\eta_2-1) + (\alpha_1-\eta_1-1|\alpha_2-\eta_2) \,, \nn
\ea
and
\ba
\frac{(\alpha_1|\alpha_2)}{(\eta_1|\alpha_2)} = (\alpha_1-\eta_1-1|0) \,\,,\hspace{1cm} \frac{(\alpha_1|\alpha_2)}{(\alpha_1|\eta_2)} = (0|\alpha_2-\eta_2-1) \,. \nn
\ea
Specializing the RHS to the arguments $x_i = q^{\rho_i}$ and plugging in (\ref{schurhook}) then yields
\ba \label{hookdecomp}
\frac{(\alpha_1|\alpha_2)}{(\eta_1|\eta_2)} &=& \frac{q^{\frac{1}{4}(\hkappa(\alpha_1-\eta_1-1)-\hkappa(\alpha_2-\eta_2-1))}}{ [\alpha_1 -\eta_1]![\alpha_2-\eta_2]!} \\
&=&  (\alpha_1-\eta_1-1|0) (0|\alpha_2-\eta_2-1) \,. \nn
\ea
In the last line, we have extended the definition of Schur polynomials by setting $(0|-1) = (-1|0) =1$.

Three identities we use repeatedly are
\ba
\alpha &=& q^{\frac{\kappa(\alpha)}{2}} \alpha^t \,, \nn\\
\alpha \,\beta(\alpha) &=& \beta \,\alpha (\beta)  \,, \nn\\
q^{\frac{\kappa(\beta)}{2}} \alpha \,\beta^t( \alpha^t) &=& \sum_\eta \frac{\alpha}{\eta} \frac{\beta}{\eta} \,.\nn
\ea
The last two identities are known to hold by explicit evaluation for many examples, and are required for the cyclicity of the vertex to hold, but we are not aware of a proof.

Using (\ref{hookdecomp}), the decomposition of the vertex with two non-trivial representations as in (\ref{2nontriv}) follows easily,
\ba
C_{\lambda \mu \cdot}  &=& \lambda \, \mu + q^{\frac{\kappa(\lambda)}{2}}\sum_{\eta \neq \cdot} \frac{\lambda^t}{\eta} \frac{\mu}{\eta} \nn\\
&=& \lambda \, \mu + q^{\frac{\kappa(\lambda_1,\lambda_2)}{2}} \sum_{\eta_1=0}^{\min(\lambda_2,\mu_1)} \sum_{\eta_2=0}^{\min(\lambda_1,\mu_2)} \frac{(\lambda_2|\lambda_1)}{(\eta_1| \eta_2)} \frac{(\mu_1|\mu_2)}{(\eta_1|\eta_2)} \nn \\
&=& \lambda \, \mu + \sum_{\eta_1=0}^{\min(\lambda_2,\mu_1)} (\lambda_2 - \eta_1-1|0) (\mu_1 - \eta_1-1|0) q^{-\frac{1}{2}\hkappa(\lambda_2)} \times \nn \\
& & \hspace{3cm} \times \sum_{\eta_2=0}^{\min(\lambda_1,\mu_2)} (0| \lambda_1-\eta_2-1)(0|\mu_2 - \eta_2-1) q^{\frac{1}{2} \hkappa(\lambda_1)}  \nn \,.
\ea

To express the partition function of the conifold, we have, following \cite{strip} introduced the notation,
\ba
\blb \alpha \beta\brb &=& \exp
\left( \sum_k C_k(\alpha, \beta) \log (1- Q q^k) -  \sum_{n=1}^{\infty} \frac{(Qq)^n}{n(1-q^n)^2} \right) \nn \\
&=& \prod_{n=1}^{\infty} (1-Qq^n)^n \prod_k (1-Qq^k)^{C_k(\alpha,\beta)} \label{productform}
\ea
where
\ba \label{defck}
\sum_k C_k(\alpha, \beta)q^k &=& \frac{q}{(q-1)^2} \left( 1 + (q-1)^2 \sum_{i=1}^{d_\alpha} q^{-i} \sum_{j=0}^{\alpha_i-1} q^j \right) \left( 1 + (q-1)^2 \sum_{i=1}^{d_\beta} q^{-i} \sum_{j=0}^{\beta_i-1} q^j \right) \nonumber \\
& & - \frac{q}{(1-q)^2}  \,,
\ea
and the product over $k$ in equation (\ref{productform}) hence runs over a finite range. This is the only other expression in this paper where we have used the notation $\alpha_i$ to denote the number of boxes in the $i^{th}$ row of the Young tableaux $\alpha$, and $d_\alpha$ to denote its number of rows. We also introduce non-doubled brackets as\ba
\{ \alpha \beta \} = \blb \alpha \beta \brb / \blb \cdot \cdot \brb \,, \nn
\ea 
where we haved divided by the closed string contribution to (\ref{productform}).

We will need the expression $\{\alpha \beta \}$ for $\alpha$, $\beta$ hook representations. To this end,
\ba
\sum_k C_k(\alpha, \beta)q^k &=& \frac{q}{(q-1)^2} \left( 1 + (q-1)^2 (q^{-1} \sum_{i=0}^{\alpha_1}q^i + \sum_{i=2}^{\alpha_2+1} q^{-i})  \right) \times \nonumber \\
 & & \times \left( 1 + (q-1)^2 (q^{-1} \sum_{i=0}^{\beta_1}q^i + \sum_{i=2}^{\beta_2+1} q^{-i})  \right) - \frac{q}{(1-q)^2} \nonumber\\
&=& \sum_{i=-\alpha_2}^{\alpha_1} q^i +  \sum_{i=-\beta_2}^{\beta_1} q^i + \frac{(q-1)^2}{q} \sum_{i=-\alpha_2}^{\alpha_1} q^i \sum_{j=-\beta_2}^{\beta_1} q^j \,.\nonumber\\ \label{fewlinesabove}
\ea
For $\alpha=\cdot$ or $\beta = \cdot$, all sums involving $\alpha$, $\beta$ respectively should be set to 0. In this case,
\ba
\{ \alpha \cdot\} &=& \prod_{i = -\alpha_2}^{\alpha_1} (1-Qq^i)  \,. \label{bracketone}
\ea
For the general case, note that the final term in (\ref{fewlinesabove}) simplifies,
\ba
\frac{(q-1)^2}{q} \sum_{i=-\alpha_2}^{\alpha_1} q^i \sum_{j=-\beta_2}^{\beta_1} q^j &=& \left( \sum_{i=-\alpha_2+1}^{\alpha_1+1}  \sum_{j=-\beta_2}^{\beta_1} + \sum_{i=-\alpha_2-1}^{\alpha_1-1}  \sum_{j=-\beta_2}^{\beta_1} - 2 \sum_{i=-\alpha_2}^{\alpha_1}  \sum_{j=-\beta_2}^{\beta_1}\right) q^{i+j} \nn \\
&=& (q^{\alpha_1+1} - q^{-\alpha_2} -q^{\alpha_1} + q^{-\alpha_2-1})\sum_{j=-\beta_2}^{\beta_1} q^j \nn\\
&=& (-q^{\alpha_1} + q^{-\alpha_2-1})(1- q)
\frac{q^{-\beta_2}-q^{\beta_1+1}}{1-q} \nn\\
&=&(-q^{\alpha_1} + q^{-\alpha_2-1})(q^{-\beta_2}-q^{\beta_1+1})\,.\nn
\ea
Hence, for $\alpha$ and $\beta$ both non-trivial,
\ba
\{\alpha \beta \} &=&  \{\alpha \cdot\} \{\beta \cdot \} \frac{(1-Qq^{\alpha_1 + \beta_1 +1}) (1-Qq^{-(\alpha_2+\beta_2+1)})}{(1-Qq^{\alpha_1 - \beta_2})(1-Q q^{-\alpha_2+\beta_1})} \,. \nn
\ea
The decomposition (\ref{2nontriv}) for configuration 1 in figure \ref{fig:fourtwos} is determined as follows,
\ba
(-1)^{|\beta|} Z_{\alpha \beta  \cdot \cdot} / \blb \cdot \cdot \brb & =&\beta^t  \{\beta \cdot \}   \sum_{\eta} \frac{\alpha}{\eta}(\beta^t) \eta^t (-Q)^{|\eta|} \nn \\
&=& \{\beta \cdot \}  \sum_{\eta,\nu} c^{\alpha}_{\eta \nu}\,  \nu(\beta^t) \beta^t \,\eta^t (-Q)^{|\eta|} \nn \\
&=& q^{-\frac{\kappa(\beta)}{2}} \{\beta \cdot \}  \sum_{\eta,\nu} c^{\alpha}_{\eta \nu}\, q^{\frac{\kappa(\nu)}{2}} \sum_\kappa \frac{\nu^t}{\kappa} \frac{\beta}{\kappa} \eta^t (-Q)^{|\eta|} \nn\\
&=&  \{\beta \cdot \} \sum_{\eta} c^{\alpha}_{\eta \nu} \nu \,\beta^t \, \eta^t (-Q)^{|\eta|} + q^{-\frac{\kappa(\beta)}{2}} \{\beta \cdot \}  \sum_{\eta,\nu} c^{\alpha}_{\eta \nu}\, q^{\frac{\kappa(\nu)}{2}} \sum_{\kappa \neq \cdot} \frac{\nu^t}{\kappa} \frac{\beta}{\kappa} \eta^t (-Q)^{|\eta|} \nn \\
&=& \beta^t \{\beta \cdot \} \sum_{\eta} \frac{\alpha}{\eta} \eta^t (-Q)^{|\eta|} +q^{-\frac{\kappa(\beta)}{2}} \{\beta \cdot \} \sum_{\eta,\nu} c^{\alpha}_{\eta \nu}\,  \eta^t (-Q)^{|\eta|} q^{\frac{\kappa(\nu)}{2}} \sum_{\kappa \neq \cdot} \frac{\nu^t}{\kappa} \frac{\beta}{\kappa}\nn \\
&=& \alpha \{\alpha^t \cdot\} \beta^t \{\beta \cdot \}+q^{-\frac{\kappa(\beta)}{2}}  \{\beta \cdot \}  \sum_{\nu} \frac{\alpha^t}{\nu^t} (-Q)^{|\alpha|-|\nu|} \, q^{\frac{\kappa(\nu)}{2}} \sum_{\kappa \neq \cdot} \frac{\nu^t}{\kappa} \frac{\beta}{\kappa}  \,.\nn \\ \label{2alongsame}
\ea
For configuration 3 in figure \ref{fig:fourtwos}, the following calculations enters,
\ba
\frac{(1-Qq^{\alpha_1+\beta_1+1})(1-Qq^{-(\alpha_2+\beta_2+1)})}{(1-Qq^{\alpha_1-\beta_2})(1-Qq^{\beta_1-\alpha_2})} - 1 &=& \frac{Q(-q^{\alpha_1+\beta_1+1} - q^{-(\alpha_2+\beta_2+1)} +q^{\alpha_1-\beta_2} + q^{\beta_1 - \alpha_2})}{(1-Qq^{\alpha_1-\beta_2})(1-Qq^{\beta_1-\alpha_2})} \nn \\
&=& -Q \frac{(q^{\alpha_1}-q^{-\alpha_2-1})(q^{\beta_1+1}-q^{-\beta_2-2}) }{(1-Qq^{\alpha_1-\beta_2})(1-Qq^{\beta_1-\alpha_2})} \nn\\
&=& -Q \frac{q^{\frac{\alpha_1-\alpha_2-1}{2}} [\alpha_1+\alpha_2+1] q^{\frac{\beta_1-\beta_2+1}{2}}[\beta_1+\beta_2+1] }{(1-Qq^{\alpha_1-\beta_2})(1-Qq^{\beta_1-\alpha_2})} \nn \,,
\ea
and
\ba
Z_{\cdot \alpha  \cdot \beta }/\blb \cdot \cdot \brb &=& (-1)^{|\alpha|} \alpha^t \{\alpha \cdot \}_Q (-1)^{|\beta|} \beta^t \{\beta \cdot \}_Q + \nn\\
& & (-1)^{\alpha_1+\beta_2} \frac{q^{\frac{1}{4} (\hkappa(\beta_2) -\hkappa(\alpha_1))}}{[\alpha_1]! [\beta_2]!}(-1)^{\alpha_2+\beta_1} \frac{q^{\frac{1}{4} (\hkappa(\alpha_2) - \hkappa(\beta_1))}}{[\alpha_2]! [\beta_1]!} \times \nn\\
& &(1-Q)^2 \prod_{i=1}^{\alpha_1} (1-Qq^i)\prod_{i=1}^{\alpha_2} (1-Qq^{-i})
\prod_{i=1}^{\beta_1} (1-Qq^i) \prod_{i=1}^{\beta_2} (1-Qq^{-i})\times \nn \\
& &
(-Q) \frac{q^{\frac{\alpha_1-\alpha_2-1}{2}}}{1-Qq^{\alpha_1-\beta_2} } \frac{q^{\frac{\beta_1-\beta_2+1}{2}}}{1-Qq^{\beta_1-\alpha_2}} \,.\label{disent}
\ea

To derive the relation
\ba 
\langle 0 | \psi^i(y) \psi^{i*}(x) | Z \rangle &=& \frac{1}{e^{y} - e^{x}} \Psi^i(y) \Psi^{i*}(x) \,,
\ea
for any $i=1, \ldots, 4$, compare
\ba
\lefteqn{(e^{y} - e^x) \langle 0 | \psi^i(y) \psi^{i*}(x) | Z \rangle =} \nn\\
& & 1 - \sum_{n=0}^{\infty} a_{0n}^{ii} e^{-(n+1)y} + \sum_{m=0}^{\infty} a_{m0}^{ii} e^{-(m+1)x} + \sum_{m,n=0}^{\infty} (a_{m \,n+1}^{ii} - a_{m+1\, n}^{ii}) e^{-(m+1)x} e^{-(n+1) y} \,. \nn
\ea
to
\ba
\Psi^i(y) \Psi^i(x)^* &=& 1 - \sum_{n=1}^{\infty} a_{0n}^{ii} e^{-(n+1)y} + \sum_{m=1}^{\infty} a_{m0}^{ii} e^{-(m+1)x} - \sum_{m,n=1}^{\infty} a_{0n}^{ii} a_{m0}^{ii} e^{-(m+1)x-(n+1)y} \nn
\ea 
The relation follows by
\ba
a_{m-1 \,n}^{ii} - a_{m\, n-1}^{ii} &=& (-1)^{k} \,\frac{q^{\frac{m(m-1)-n(n-1)}{4}}}{[m]![n]!} \prod_{j=-(m-1)}^{n-1} (1-Qq^j) (1-Q)  \nn \\
&=& - a^{ii}_{0\,n-1} a^{ii}_{m-1\,0} \,, \nn
\ea
with $k=m$ for $i=1,3$, and $k=n$ for $i=2,4$, using
\ba
q^{\frac{m}{2}} (1-Qq^{-m}) [n] + q^{-\frac{n}{2}} (1-Qq^n)[m] &=& (1-Q)[m+n]  \,. \nn
\ea

Finally, we derive the relation between $a^{24}$ and $a^{22}, a^{44}$ announced in section \ref{s:transop}. Consider $(-1)^{\beta_2} a^{42}_{\alpha_1,\beta_2}$,
\ba
\lefteqn{C^{42}\sum_{\eta_1=0}^{\alpha_1} \sum_{\nu_2=0}^{\beta_2} \sum_{\kappa=0}^{\min(\eta_1,\nu_2)} (\alpha_1 - \eta_1-1|0)(0|\eta_1-\kappa-1) (0|\beta_2-\nu_2-1) (\nu_2 - \kappa-1|0)(-Q)^{\eta_1+\nu_2-\kappa}=} \nn\\
&=& C^{42} \sum_{\kappa=0}^{\min(\alpha_1,\beta_2)} \sum_{\mu=0}^{\alpha_1-\kappa} \sum_{\nu=0}^{\beta_2-\kappa} (\alpha_1 - \kappa - \mu-1|0) (0| \mu-1) (0|\beta_2 - \kappa - \nu-1)(\nu-1|0) (-Q)^{\mu + \nu + \kappa} \nn\\
&=& C^{42} \sum_{\kappa=0}^{\min(\alpha_1,\beta_2)} \frac{(\alpha_1-\kappa-1|0)}{\mu} \mu^t (-Q)^{|\mu|} \frac{(0|\beta_2-\kappa-1)}{\nu} \nu^t (-Q)^{|\nu|} (-Q)^{\kappa} \nn\\
&=& C^{42} \sum_{\kappa=0}^{\min(\alpha_1,\beta_2)} (\alpha_1-\kappa-1|0) \,\{(0|\alpha_1 - \kappa-1) \cdot \} \,(0|\beta_2-\kappa-1) \,\{ (\beta_2 - \kappa-1|0) \cdot \} \,(-Q)^{\kappa}  \nn\\
&=& C^{42} \sum_{\kappa=0}^{\min(\alpha_1,\beta_2)} (-1)^{\beta_2-\kappa-1}a^{22}_{\alpha_1 - \kappa -1,0} \,a^{44}_{0,\beta_2 - \kappa -1} (-Q)^{\kappa}  \,.\nn
\ea
Note that we use $\mu$ and $\nu$ to denote first the number of rows, columns respectively of a hook representation, and then a general representation.

\end{document}